\documentclass[reprint,nofootinbib,aps,prd,superscriptaddress,floatfix,amsmath,amssymb]{revtex4-2}

\usepackage{graphicx}
\usepackage{dcolumn}% Align table columns on decimal point
\usepackage{bm}% bold math
\usepackage{braket}
\usepackage{breqn}
\usepackage{color}

\usepackage[colorlinks=true,linkcolor=red,citecolor=blue]{hyperref}

\begin{document}

%\preprint{}

\title{Evolution of binary systems accompanying axion clouds \\
in extreme mass ratio inspirals}



\author{Takuya Takahashi}
\email{t.takahashi@tap.scphys.kyoto-u.ac.jp}
\affiliation{Department of Physics$,$ Kyoto University$,$ Kyoto 606-8502$,$ Japan}
\author{Hidetoshi Omiya}
\email{omiya@tap.scphys.kyoto-u.ac.jp}
\affiliation{Department of Physics$,$ Kyoto University$,$ Kyoto 606-8502$,$ Japan}
\author{Takahiro Tanaka}
\email{t.tanaka@tap.scphys.kyoto-u.ac.jp}
\affiliation{Department of Physics$,$ Kyoto University$,$ Kyoto 606-8502$,$ Japan}
\affiliation{Center for Gravitational Physics and Qunatum Information$,$ Yukawa Institute for Theoretical Physics$,$ Kyoto University$,$ Kyoto 606-8502$,$ Japan}


\date{\today}

\begin{abstract}
Superradiant instability of rotating black holes (BHs) leads to the formation of a cloud of ultralight bosons, such as axions.
When the BH with the cloud belongs to a binary system and is in an inspiraling orbit, the resonant transition between the axion's bound states can occur.
We study the history of the evolution of the binary system accompanying the cloud composed of the fastest growing mode, and its impact on the observational signatures, especially for small mass ratio cases.
In this case, the hyperfine resonance, which has a very small resonance frequency, is relevant.
Therefore, due to the long timescale, we should take into account the decaying process of axions in the transition destination mode, the backreaction to the orbital motion and the central BH, and gravitational emission from the cloud.
We present a formulation to examine the evolution of the system around the resonance and useful expressions for the analysis.
As a result, we found the mass of the cloud that can remain after the resonance is, at most, about $10^{-5}$ of the central BH. The maximum remaining cloud mass is achieved when the mass ratio of the binary is $q\sim10^{-3}$. 
In addition, we show that the resonant transition hardly changes the BH mass and spin distribution, while the associated modification of the gravitational wave frequency evolution when the binary pass through the resonance can be a signature of the presence of the cloud. 
\end{abstract}

%\key words{}

\maketitle

\section{Introduction}\label{Sec:intro}
Ultralight bosons, such as axions or axion-like particles, can cause various phenomena in the universe.
Such particles are universally predicted by string theory~\cite{Arvanitaki:2009fg,Svrcek:2006yi} and can be a candidate for dark matter~\cite{Dine:1982ah,Preskill:1982cy,Abbott:1982af,Hui:2016ltb}.
They can be weakly coupled to the Standard Model particles, but even in such a case 
the gravitational interaction with black holes (BHs) 
and related gravitational waves (GWs) can provide a new avenue to explore them observationally.

The existence of massive bosonic fields induces the superradiant instability around rotating BHs~\cite{Press:1972zz,Brito:2020oca}.
Bosons with mass in the range $10^{-20}\sim 10^{-10}$ eV have the Compton wavelength 
comparable to the size of astrophysical BHs, and extract energy and angular momentum efficiently to form a condensate~\cite{Detweiler:1980uk,Dolan:2007mj}.
We refer to the condensate as an axion cloud and the composing particles simply as axions.
The cloud formation makes astrophysical observable imprints, such as a forbidden region in the distribution of mass and spin of BHs~\cite{Arvanitaki:2010sy,Brito:2014wla,Stott:2018opm,Du:2022trq} and continuous GW emission~\cite{Arvanitaki:2014wva,Arvanitaki:2016qwi,Yoshino:2013ofa,Brito:2017zvb,Isi:2018pzk,Ng:2020jqd,Siemonsen:2022yyf}.

In this paper, we focus on the cases where BHs with clouds belong to binary systems.
GWs from the binary inspiral can be a signature to examine the environment around BHs including the cloud~\cite{Baryakhtar:2022hbu,Bamber:2022pbs,Cole:2022fir,DeLuca:2021ite,DeLuca:2022xlz,Kim:2022mdj}.
Axion clouds occupy a quasi-bound state of axions, which is usually the fastest growing mode.
During the inspiral phase, the tidal interaction from the companion acts as an oscillating tidal field.
It induces the resonant transition to another mode when the orbital frequency coincides with the phase velocity 
difference between the original mode of the cloud and the other~\cite{Baumann:2018vus,Baumann:2019ztm}.
The change of the orbital motion of the binary and the associated GW frequency due to the backreaction can also be a signature of the presence of the cloud~\cite{Baumann:2019ztm,Ding:2020bnl,Tong:2021whq,Baumann:2022pkl}.
To clarify the impact on the observational signatures, it is important to understand the history of the evolution during the inspiral phase.

If the separation of the binary is sufficiently small, the cloud configuration is tidally disrupted~\cite{Takahashi:2021yhy,Cardoso:2020hca}, 
and the transition to unbound states occurs~\cite{Takahashi:2021yhy,Baumann:2021fkf}.
However, for binary systems formed with a sufficiently large separation, 
the resonant transition should first occur with the smallest possible resonance frequency.
The frequency spectrum of axion eigenmodes possesses the structure of hyperfine splittings due to the rotation of the central BH~\cite{Baumann:2019eav}, 
and the resonance frequency associated with the hyperfine splitting is the smallest one.
In Ref.~\cite{Takahashi:2021yhy}, we showed that, for nearly equal mass binaries, 
this hyperfine resonance can be neglected since the resonance condition is not maintained long enough because of 
the decrease of the angular momentum of the cloud itself. 
We also showed that, before the transition caused by the leading quadrupole moment  of the tidal potential occurs, the cloud is disrupted by the effects of higher multipole moments, 
and finally the cloud is depleted as a result of transitions to unbound states. 

In contrast to nearly equal mass binaries, for small mass ratio binaries, the hyperfine resonance should be considered 
because of a large backreaction to the orbital motion, which maintains the orbital frequency within the resonance band 
for a long period.
It has great importance to examine the dynamics of small mass ratio binaries, because they are one of the main targets for future GW observations, such as LISA~\cite{Audley:2017drz}.
In this case, because of the very long timescale of the binary evolution due to the radiation reaction, 
some effects that can be neglected for the transition for nearly equal mass binaries become relevant.

First, the decay of non-superradiant transition destination modes and the backreaction to the central BH mass and spin become relevant.
Since the resonance band is broadened corresponding to the imaginary part of the frequencies of decaying destination modes, 
the transition timescale staying within the resonance band becomes even longer. 
Therefore, we should also take into account the GW emission from the cloud during the transition.
We develop a formulation that includes all of these effects within the adiabatic approximation.
It is difficult to solve the originally obtained set of equations throughout the whole period across 
the resonance band, since the solution oscillates rapidly. 
To overcome this difficulty, we also present a method to give an approximate solution with sufficient accuracy.

In this paper, we consider axion clouds in a non-relativistic regime, and neglect the self-interaction of axions, for simplicity. 
For a relativistic regime, the energy spectrum deviates significantly from the one obtained by non-relativistic approximation, and the transition to be considered can change~\cite{Berti:2019wnn,Takahashi:2021eso}.
In addition, the self-interaction can play an important role during the formation of the cloud~\cite{Gruzinov:2016hcq,Fukuda:2019ewf,Baryakhtar:2020gao,Omiya:2020vji,Omiya:2022mwv,Omiya:2022gwu, Branco:2023frw,Chia:2022udn}.
Here, we leave considering these effects as future work, to focus on the tidal effect in binary systems.

This paper is organized as follows. In Sec.~\ref{Sec:elements}, we review the elements involved in the evolution of axion clouds in binary systems.  In Sec.~\ref{Sec:formulation}, we present a formulation for examining the hyperfine resonance in small mass ratio binaries. In Sec.~\ref{Sec:results}, we discuss the results obtained using our formulation. Finally, we give a summary and conclusion in Sec.~\ref{Sec:summary}. Throughout this paper, we use the unit with $c=\hbar=G=1$.

\section{Elements involved in the evolution of axion clouds}\label{Sec:elements}

In this section, we summarize the elements involved in describing the evolution of axion clouds, especially during the binary inspirals. Consider a scalar field (axion) of mass $\mu$ around a rotating BH belonging to a binary system. We denote the central BH mass by $M$ and angular momentum by $J=aM=\chi M^2$. Formally, we can write the equation of motion for axion on a spacetime with the metric  $\tilde{g}_{\mu\nu}=g_{\mu\nu}+h_{\mu\nu}$ as
\begin{equation}\label{eom}
(\tilde{g}^{\mu\nu}\tilde{\nabla}_{\mu}\tilde{\nabla}_{\nu}-\mu^2)\phi=0~,
\end{equation}
where $g_{\mu\nu}$ is the Kerr metric. 
We consider the tidal field from the binary companion and the decay due to the gravitational wave emission from the cloud as contributions to the perturbation. 
As we will see later, since there is a hierarchy of frequencies between them, we can treat them separately. 
We first review the features of axion clouds in the unperturbed background, 
and later the effects of the tidal interaction and the GW emission.

\subsection{Energy spectrum and superradiance}\label{sub:background}
In the non-relativistic regime, it is appropriate to introduce a new complex 
scalar field variable $\psi$ by 
\begin{align}
\phi=\frac{1}{\sqrt{2\mu}}\left(e^{-i\mu t}\psi+e^{i\mu t}\psi^{\ast}\right)~. 
\end{align}
We assume that $\psi$ changes slowly in time compared to the timescale determined by $\mu^{-1}$. 
Then, we can ignore the $\partial_t^2\psi$ term and rewrite the background equation of motion \eqref{eom} as
\begin{align}
i\frac{\partial}{\partial t}\psi=H_0\psi~, \quad H_0=-\frac{1}{2\mu}\nabla^2-\frac{\alpha}{r}+{\cal O}(\alpha^2)~, 
\end{align}
where we have introduced the gravitational fine structure constant $\alpha\equiv M\mu$, 
and this approximation is well justified for $\alpha\ll 1$.
Solving this equation with the ingoing boundary condition at the BH horizon 
and the exponentially decaying boundary condition at infinity, 
we have the quasi-bound eigenstate $\varphi_{nlm}(\bm{r})$ that satisfies 
$H_0\varphi_{nlm}=(\omega_{nlm}-\mu)\varphi_{nlm}$. 
They are labeled by the principal, azimuthal and magnetic quantum numbers 
like a hydrogen atom. 
The eigenfrequency is approximately given by
\begin{equation}
\omega_{nlm}=(\omega_{R})_{nlm}+i(\omega_{I})_{nlm}~,
\end{equation}
with~\cite{Baumann:2019eav,Baumann:2018vus} 
\begin{align}
(\omega_{R})_{nlm}=\mu\left(1-\frac{\alpha^2}{2n^2}-\frac{\alpha^4}{8n^4}+\frac{(2l-3n+1)\alpha^4}{n^4(l+1/2)} \right.& \notag \\
\left.+\frac{2m\chi\alpha^5}{n^3 l(l+1/2)(l+1)}\right)& ~, \label{Ene} 
\end{align}
\begin{align}
(\omega_{I})_{nlm}=2(r_{+}/M)C_{nlm}(a,\alpha)(m\Omega_{H}-\omega_{nlm})\alpha^{4l+5}~, \label{omeI}
\end{align}
where $r_{+}=M+\sqrt{M^2-a^2}$ is the horizon radius, 
$\Omega_{H}=a/2Mr_{+}$ is the angular velocity of the BH horizon 
and the explicit form of $C_{nlm}(a,\alpha)$ can be found in Ref.~\cite{Baumann:2018vus}\nobreak
\footnote{It was first derived in Ref.~\cite{Detweiler:1980uk}, and corrected by a factor of $1/2$ \cite{Pani:2012bp,Bao:2022hew}.}.

As one can see from Eq.~\eqref{omeI}, 
the eigenfrequency of a mode satisfying $\omega_{R}<m\Omega_{H}$ has a positive imaginary part, 
and the cloud grows exponentially by the superradiance. 
The mode $\ket{nlm}=\ket{211}$ is the fastest growing mode for $\alpha\lesssim0.45$.
The BH spin decreases as the cloud grows until the superradiance condition is saturated.
The critical spin at which the superradiance terminates is approximately given by
\begin{align}\label{eq:acrit}
\chi_{\rm crit}=\frac{4m\alpha}{m^2+4\alpha^2}~.
\end{align}

The real part of the eigenfrequency can be regarded as eigenenergy, and its degeneracy among the modes with only $m$ being different is solved due to the rotation of the BH at the order of ${\cal O}(\alpha^5)$, which is called the ``hyperfine" splitting.

\subsection{Tidal interaction}\label{sub:tidal}
When a BH accompanied by an axion cloud belongs to a binary system, 
the tidal field from the companion introduces a perturbation. 
The general state of the cloud can be expressed by 
\begin{align}
\psi=\sum_i c_i(t)\varphi_i\,,
\end{align}
as a superposition of 
orthonormal eigenfunctions $\varphi_i$. 
Under the same approximation taken in the preceding subsection, the equation of motion with the perturbation is given by
\begin{align}\label{eom:pert}
i\frac{dc_i}{dt}=\sum_j \left((\omega_j-\mu)\delta_{ij}+\int d^3x\ \varphi^{\ast}_i V_{\ast}\varphi_j\right)c_j~.
\end{align}
For simplicity, we assume that the binary orbit is quasi-circular and on the plane perpendicular to the central BH spin. By multipole expansion, we can write the tidal field from the companion of mass $M_{\ast}$ at $\bm{r}(t)=(R_{\ast}(t),\Theta_{\ast}(=\pi/2),\Phi_{\ast}(t))$ as
\begin{align}
V_{\ast}&=\frac{1}{2}\mu h^{tt}_{\rm tidal} \notag \\
&=-q\alpha\sum_{l_{\ast}m_{\ast}}\frac{4\pi}{2l_{\ast}+1}\frac{r^{l_{\ast}}_{<}}{r^{l_{\ast}+1}_{>}}Y_{l_{\ast}m_{\ast}}^{\ast}(\Theta_{\ast},\Phi_{\ast})Y_{l_{\ast}m_{\ast}}(\theta,\phi)~,
\end{align}
where $q\equiv M_{\ast}/M$ is the mass ratio, $r_{>}(r_{<})$ is the larger (smaller) of $r$ and $R_{\ast}$, and $Y_{lm}$ are the spherical harmonics. The angular velocity of the binary is defined by $\dot{\Phi}_{\ast}(t)=\pm\Omega(t)$, and the upper (lower) sign represents the case of co-rotating (counter-rotating) orbits. 
Since this interaction oscillates quasi-periodically, it works efficiently only when the orbital angular velocity is close to the difference between the phase velocity of the two modes. Therefore, it is sufficient to consider a two-mode subspace~\cite{Baumann:2019ztm}.
Tidal field mixes two modes, and the time evolution of particle number in each mode is, from Eq.\eqref{eom:pert}, given by 
\begin{align}
i\dot{\bm c}={\cal H}{\bm c}
\end{align}
with
\begin{align}\label{Sch1}
{\cal H}=\begin{pmatrix}
-\Delta E/2+i\omega_{I}^{(1)} & \eta e^{i\Delta m\Phi_{\ast}} \\
\eta e^{-i\Delta m\Phi_{\ast}} & \Delta E/2+i\omega_{I}^{(2)}
\end{pmatrix}~,
\end{align}
where $\Delta E=\omega_{R}^{(2)}-\omega_{R}^{(1)}$, $\Delta m=m_2-m_1$, and $\eta(t)=\left|\int d^3x\ \varphi^{\ast}_2 V_{\ast}\varphi_1\right|$. To remove the rapidly oscillating term, we perform the unitary transformation as $c\to{\cal U}^{-1}c$ and ${\cal H}\to{\cal U}^{\dagger}{\cal H}\,{\cal U}-i\,{\cal U}^{\dagger}{\cal \dot{U}}$ with the matrix ${\cal U}(t)={\rm diag}(e^{i\Delta m\Phi_{\ast}/2},e^{-i\Delta m\Phi_{\ast}/2})$.
As a result, we can describe the level transition due to the tidal field by
\begin{align}\label{Sch2}
{\cal H}=\begin{pmatrix}
\pm\frac{\Delta m}{2}(\Omega-\Omega_{\rm res})+i\omega_{I}^{(1)} & \eta \\
\eta & \mp\frac{\Delta m}{2}(\Omega-\Omega_{\rm res}
)+i\omega_{I}^{(2)}
\end{pmatrix}~,
\end{align}
where we defined the ``resonance" frequency by $\Omega_{\rm res}=\pm\Delta E/\Delta m$. Now, we are interested in the time evolution of the occupation number of each state, $|c_i(t)|^2$.

\subsection{Gravitational wave emission}\label{sub:gw}
After an axion cloud forms, it dissipates through the emission of GWs. 
Here, we assume that the cloud is composed of a single mode as $\psi=c_1\varphi_1$. 
In this case, we can neglect the GW emission due to the spontaneous level transition, and GWs are sourced by the pair-annihilation of axions. 
The frequency of GWs is given by $\omega_{\rm GW} = 2 \omega_{R} \sim 2\mu$.
The energy flux of GWs from the $l=m=1$ cloud is given by~\cite{Yoshino:2013ofa}
\begin{align}\label{GWflux}
\frac{d{E}_{\rm GW}}{dt}=C\left(\frac{M_{\rm c}}{M}\right)^2\alpha^{14}~,
\end{align}
where $C$ is a numerical factor. In our analysis, we adopt $C=(484+9\pi^2)/23040$ calculated in Ref.~\cite{Brito:2014wla}. Here, $M_{\rm c}$ is the mass of the cloud defined by $M_{\rm c}=-\int d^3x\ {T^t}_t$, where ${T^t}_t$ is the $t$-$t$ component of the energy momentum tensor. According to this, the wave function $\psi$ is normalized as $|c_1|^2(=\int d^3x|\psi|^2)=M_{\rm c}/\mu$ at the leading order in $\alpha$.

When we consider only the effect of GW emission, energy conservation implies that $\dot{M}_{\rm c}=-\dot{E}_{\rm GW}$. We set the initial mass of the cloud to $M_{\rm c,0}$ at $t=t_0$. Here, we define the normalized particle number by $n_1(t)=\mu|c_1(t)|^2/M_{\rm c,0}$, and write $M_{\rm c}(t)=M_{\rm c,0}n_1(t)$.
Energy conservation reads
\begin{align}\label{eq:GWn}
\frac{dn_1}{dt}=-\frac{C}{M}\left(\frac{M_{\rm c,0}}{M}\right) n_1^2\alpha^{14}~.
\end{align}

\section{Formulation}\label{Sec:formulation}
In this section, we first explain the setup of the problem that we consider and then give a formulation to investigate it.

\subsection{Setup}\label{sub:setup}

\begin{figure}
\includegraphics[scale=0.6]{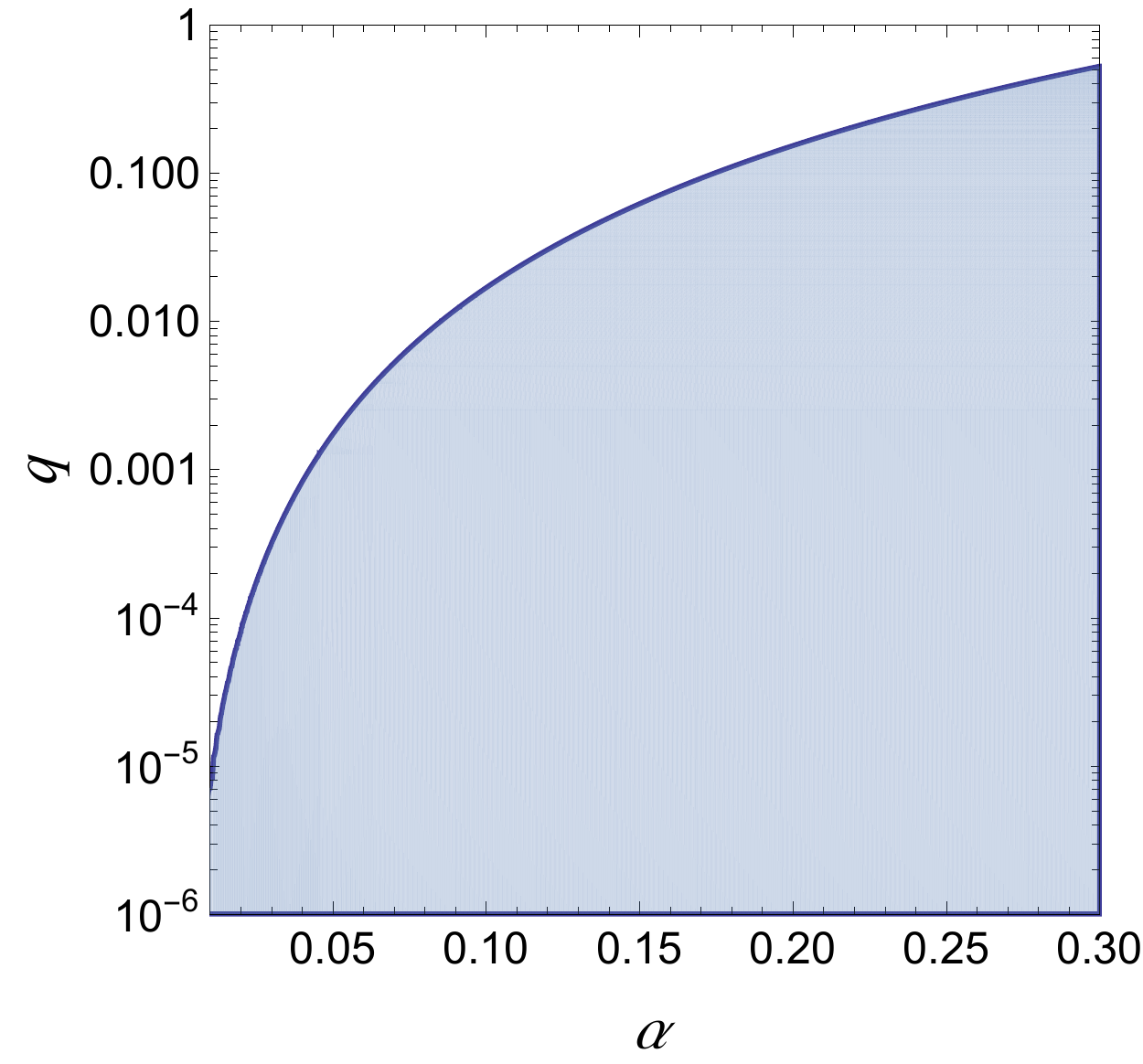}
\caption{\label{fig:region} Parameter region where the hyperfine resonance is relevant to dissipate the cloud. In the shaded region, the resonance sustains longer because the effect of the backreaction to the orbital motion is stronger than the effect of the reduction of the hyperfine splitting. The initial angular momentum of the cloud is set to $J_{\rm c,0}\to 0$. See Ref.~\cite{Takahashi:2021yhy} for the detail.}
\end{figure}

We focus on the fastest growing mode $\ket{nlm}=\ket{211}$.
We consider the situation in which the cloud is initially composed of the single mode $\ket{211}$, and the hyperfine level transition between $\ket{211}$ and $\ket{21-1}$ subsequently occurs.
Note that this transition occurs only for co-rotating orbit.
In Ref.~\cite{Takahashi:2021yhy}, we found that when the binary mass ratio $q$ is not too small, this transition does not significantly contribute to the dissipation of the cloud because of the reduction of the hyperfine splitting associated with the transfer of the angular momentum of the cloud to the orbital motion. 
However, when the mass ratio is somewhat small, the resonant tidal interaction at this hyperfine splitting frequency would largely affect the dynamics of the system. We show the parameter region where we should consider the hyperfine resonance as a process that contributes to the cloud dissipation in Fig.~\ref{fig:region}.
We investigate the latter case.

For the transition between $\ket{211}$ and $\ket{21-1}$, from Eq.\eqref{Ene}, the resonance frequency is given by\nobreak
\footnote{We do not include the contribution from the angular momentum of the cloud itself, focusing on the case where it is negligible.}
\begin{align}
\Omega_{\rm res}=\frac{\mu}{12}\chi\alpha^5~.
\end{align}
This is smaller by a factor of $\alpha^3$ than that of the ``Bohr" transition between modes with different values of $n$.
When we study the Bohr transition, $\omega_{I}$ in Eq.\eqref{Sch2} and GW flux are so small in the timescale for passing through the resonance band that we can usually neglect them\nobreak\footnote{When we consider a higher $l$ mode, the transition to the mode with smaller $l$ is allowed by the selection rule. In that case, the decay rate of the second mode can be large, and it would be important~\cite{Tong:2022bbl}.}.
However, for hyperfine transition, binary evolution around the resonance frequency is very slow and the timescale for passing through the resonance band can be large, especially for $q\ll1$.
In addition, since the angular momentum of the cloud is transferred to the orbital motion, the timescale becomes even larger. 
As a result, we should take into account not only the backreaction to the orbital motion, but also the backreaction to the mass and spin of the central BH and the effect of the GW emission from the cloud.
We summarize the timescales involved in the current problem in Appendix~\ref{app:timescale}.

In the following, we label the quantities associated with the mode $\ket{211}$ by 1, and those with $\ket{21-1}$ by 2.
For these modes, the imaginary parts of the eigenfrequencies are given by
\begin{align}
\omega_{I}^{(i)}=\frac{1}{24}\frac{r_{+}}{M}\left\{\left(1-\chi^2\right)+4r_{+}^2(m_i\Omega_{H}-\omega_{R})^2\right\}& \notag \\
\times(m_i\Omega_{H}-\omega_{R})\alpha^9& ~,
\end{align}
where $i$ is 1 or 2, and $m_1=1$ and $m_2=-1$ represent the magnetic quantum number. 
The mixing term in the Hamiltonian~\eqref{Sch2} is given by
\begin{align}\label{eq:eta}
\eta=9.0\ \frac{q}{1+q}\frac{M\Omega^2}{\alpha^3}~.
\end{align}

\subsection{Evolution of the system}\label{sub:evolution}
The dynamical timescale of the cloud can be estimated by $\omega_{R}^{-1}\simeq\mu^{-1}$. It is always short compared to the growth/decay rate of the cloud, {\it i.e.}, $(\omega_{I}^{(i)})^{-1}\gg\mu^{-1}$. 
Thus, we describe the evolution of the cloud and the central BH within the adiabatic approximation. 
The local energy and angular momentum conservation at the BH horizon reads
\begin{align}
\frac{dM}{dt}+2\omega_{I}^{(1)}M_{\rm c}^{(1)}+2\omega_{I}^{(2)}M_{\rm c}^{(2)}=0~, \label{eq:EFbalance} \\
\frac{dJ}{dt}+\frac{2\omega_{I}^{(1)}}{\mu}M_{\rm c}^{(1)}-\frac{2\omega_{I}^{(2)}}{\mu}M_{\rm c}^{(2)}= 0~, \label{eq:AMFbalance}
\end{align}
with $M_{\rm c}^{(i)}=M_{\rm c,0}n_i(t)$. Here, we used the relation between the energy flux and the angular momentum flux for each mode $\dot{J}_{\rm c}^{(i)}=(m_i/\omega_{R}^{(i)})\dot{E}_{\rm c}^{(i)}$ and the approximation $\omega_{R}=\mu$.
We denote the initial mass and angular momentum of the BH just before entering the resonance band by $M_0$ and $J_0$, and accordingly $\alpha_0=M_0\mu$.

Next, we consider the evolution of the binary system at the leading post-Newtonian order.
In clean binary systems, angular momentum conservation implies $\dot{J}_{\rm orb}=-{\cal T}_{\rm GW}$, where $J_{\rm orb}=q(1+q)^{-1/3}M_0^{5/3}\Omega^{-1/3}$ is the orbital angular momentum
and ${\cal T}_{\rm GW}$ is the torque caused by the radiation reaction due to the GW emission.
It can be rewritten as~\cite{PhysRev.136.B1224,Blanchet:2013haa}
\begin{align}
&\frac{d\Omega}{dt}=\gamma\left(\frac{\Omega}{\Omega_0}\right)^{11/3}~, \\
&\frac{\gamma}{\Omega_0^2}=\frac{96}{5}\frac{q}{(1+q)^{1/3}}(M_0\Omega_0)^{5/3}~, \label{def:gamma}
\end{align}
where the reference frequency is chosen as $\Omega_0=(\mu/12)(J_0/M_0^2)\alpha_0^5$ (which is the ``initial" resonance frequency).
Here, we add the cloud and the BH contributions to the total angular momentum conservation as $\dot{J}_{\rm orb}+\dot{J}+\dot{J}_{\rm c}^{(1)}+\dot{J}_{\rm c}^{(2)}+\dot{J}_{\rm GW}=-{\cal T}_{\rm GW}$, where $\dot{J}_{\rm GW}=(1/\mu)\dot{E}_{\rm GW}$ is the angular momentum flux of the GW from the cloud in Eq.~\eqref{GWflux}. Note that we consider GW emission only from the first mode $\ket{211}$.
(As we will see later, the particle number occupying the second mode, which is non-superradiant, 
is always tiny and does not contribute to the GW emission.) Then, we obtain\nobreak
\footnote{Strictly speaking, we should take the mass of the one paired with the companion as $M+M_{\rm c}$. However, since the cloud mass is small compared to the central BH mass, we approximated it as $M_0$.}
\begin{align}
\frac{d\Omega}{dt}=&\gamma\left(\frac{\Omega}{\Omega_0}\right)^{11/3}+R\left(\frac{\Omega}{\Omega_0}\right)^{4/3}\frac{\Omega_{0}}{M_0^{2}} \notag \\
&\times\left[\frac{d}{dt}\left(J+J_{\rm c}^{(1)}+J_{\rm c}^{(2)}\right)+\frac{1}{\mu}\frac{d{E}_{\rm GW}}{dt}\right]~, \label{eq:TotalAMC}
\end{align}
with $R=3(1+q)^{1/3}q^{-1}(M_0\Omega_0)^{1/3}$.
We take $\Omega(t_0)=\Omega_0(1+(8/3)(\gamma/\Omega_0)|t_0|)^{-3/8}$ as the initial value so that $\Omega=\Omega_0$ at $t=0$ when there are no clouds.

Finally, we describe the level transition between two modes. 
It is described by the Schr$\ddot{\rm o}$dinger equation with the Hamiltonian \eqref{Sch2}.
Note that the particle number occupying the first mode decreases due to the GW emission by pair annihilation.
Since the frequency of the emitted GW ($\omega_{\rm GW}\sim 2\mu$) is much larger than that of the tidal field ($\Omega_{\rm res}\sim\mu\alpha^6$), we can treat them separately.
Thus, we add the effect of the GW emission into the Schr$\ddot{\rm o}$dinger equation as
\begin{align}
i\frac{dc_{1}}{dt}&=\left(-(\Omega-\Omega_{\rm res})+i\omega_{I}^{(1)}-i\Gamma_{\rm GW}\right)c_{1}+\eta c_{2}~, \label{eq:LTc1} \\
i\frac{dc_{2}}{dt}&=\eta c_{1}+\left((\Omega-\Omega_{\rm res})+i\omega_{I}^{(2)}\right)c_{2}~, \label{eq:LTc2}
\end{align}
where $|c_i(t)|^2=M_{\rm c,0}n_i(t)/\mu$.
Here, $\Gamma_{\rm GW}$ represents the decay rate through the GW emission, 
whose explicit expression does not become necessary below.

From the above, the variables in this problem are $\{M,J,\Omega,c_1,c_2\}$, and we should solve the Eqs.~\eqref{eq:EFbalance}, ~\eqref{eq:AMFbalance},~\eqref{eq:TotalAMC},~\eqref{eq:LTc1}, and \eqref{eq:LTc2}. However, because of the highly oscillatory behavior of the solutions for Eqs.~\eqref{eq:LTc1} and \eqref{eq:LTc2}, it is difficult to solve these equations for a long time with sufficient accuracy.
To overcome this difficulty, we derive a set of approximate equations that can be solved easily.

\subsection{Adiabatic elimination}\label{sub:elimination}
Here, we take advantage of the fact that the decay rate of the second mode $|\omega_{I}^{(2)}|$ is large compared to the transition rate due to the mixing term $\eta$ around the resonance frequency.
Indeed, their ratio is estimated as\nobreak
\footnote{Here, we approximate $|\omega_{I}^{(2)}|\simeq \frac{1}{48}\mu\chi\alpha^{8}$ and $\chi=\chi_{\rm crit}\simeq 4\alpha$.}
\begin{align}\label{eq:decay_mixing}
\frac{|\omega_{I}^{(2)}|}{\eta}\sim 8\times 10^2\left(\frac{10^{-3}}{q}\right)\left(\frac{0.1}{\alpha}\right)~.
\end{align}
In this case, we can carry out an adiabatic elimination of the second mode and discuss with only the particle number of the first mode.

\begin{figure*}[t]
\includegraphics[keepaspectratio, scale=0.5]{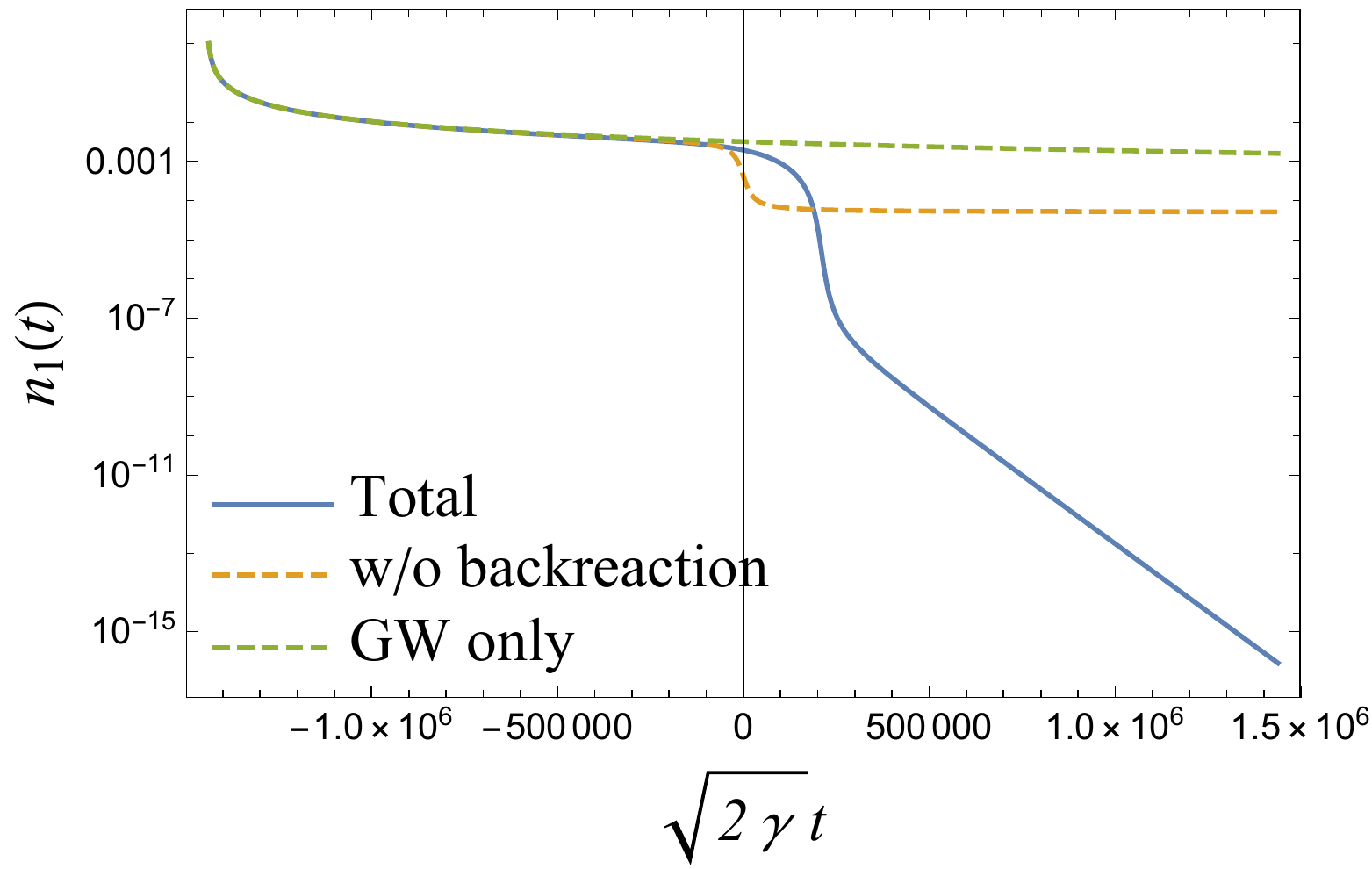}
\includegraphics[keepaspectratio, scale=0.5]{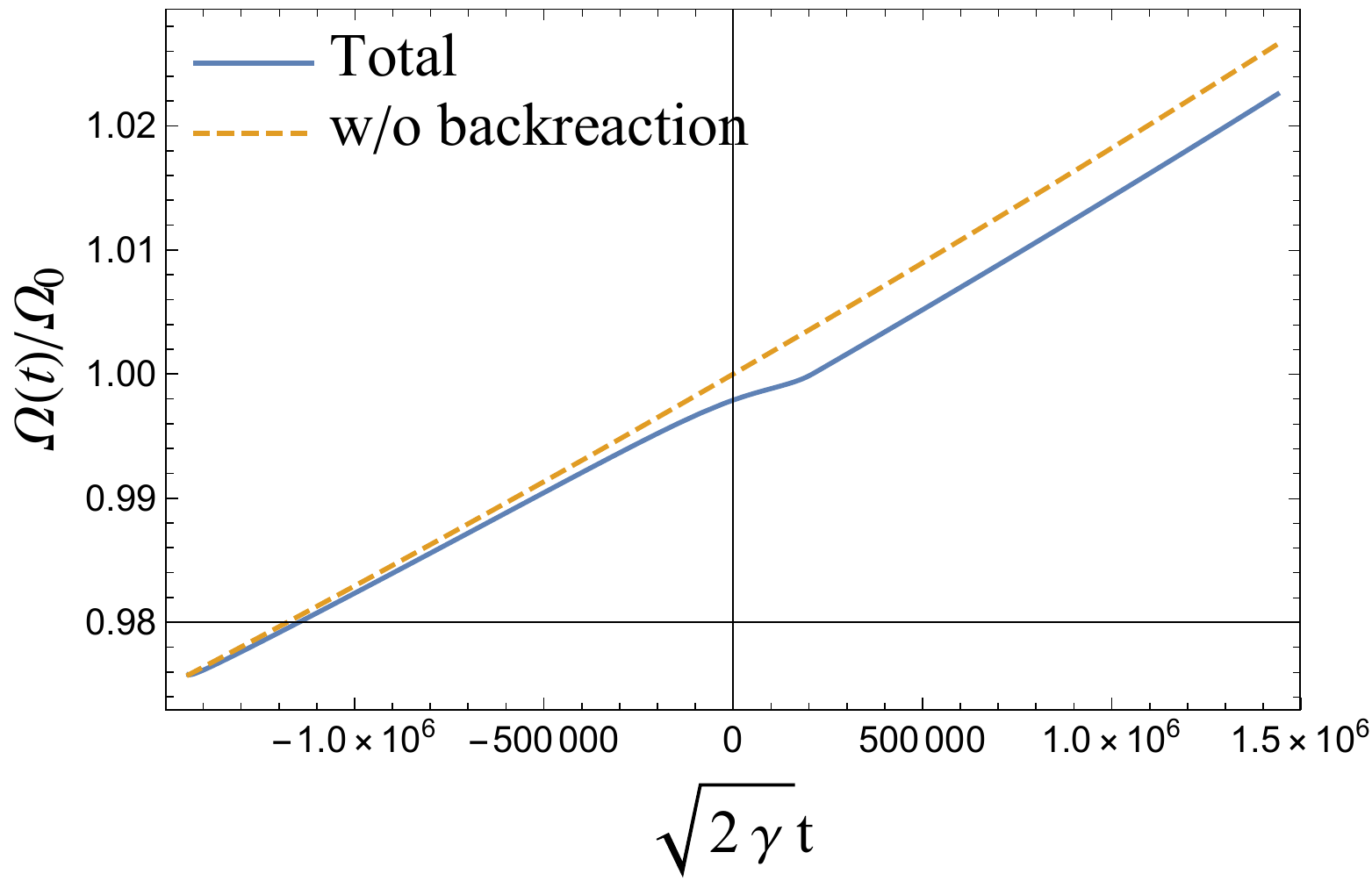}
\caption{\label{fig:n1_and_omega} Evolution of the normalized particle number of the first mode $n_1(t)$ (left) and the orbital frequency $\Omega(t)$ (right) around the resonance frequency for $q=10^{-4}$, $\alpha_0=0.1$ and $M_{\rm c,0}=10^{-3}M_0$. Blue solid lines show the results of solving all equations, and orange dashed lines show the results without taking into account the backreaction to the orbital motion and the mass and spin of the central BH. Green dashed line, in the left panel, shows the evolution of $n_1(t)$ considering only the effect of the GW emission.}
\end{figure*}

\begin{figure*}[t]
\includegraphics[keepaspectratio, scale=0.4]{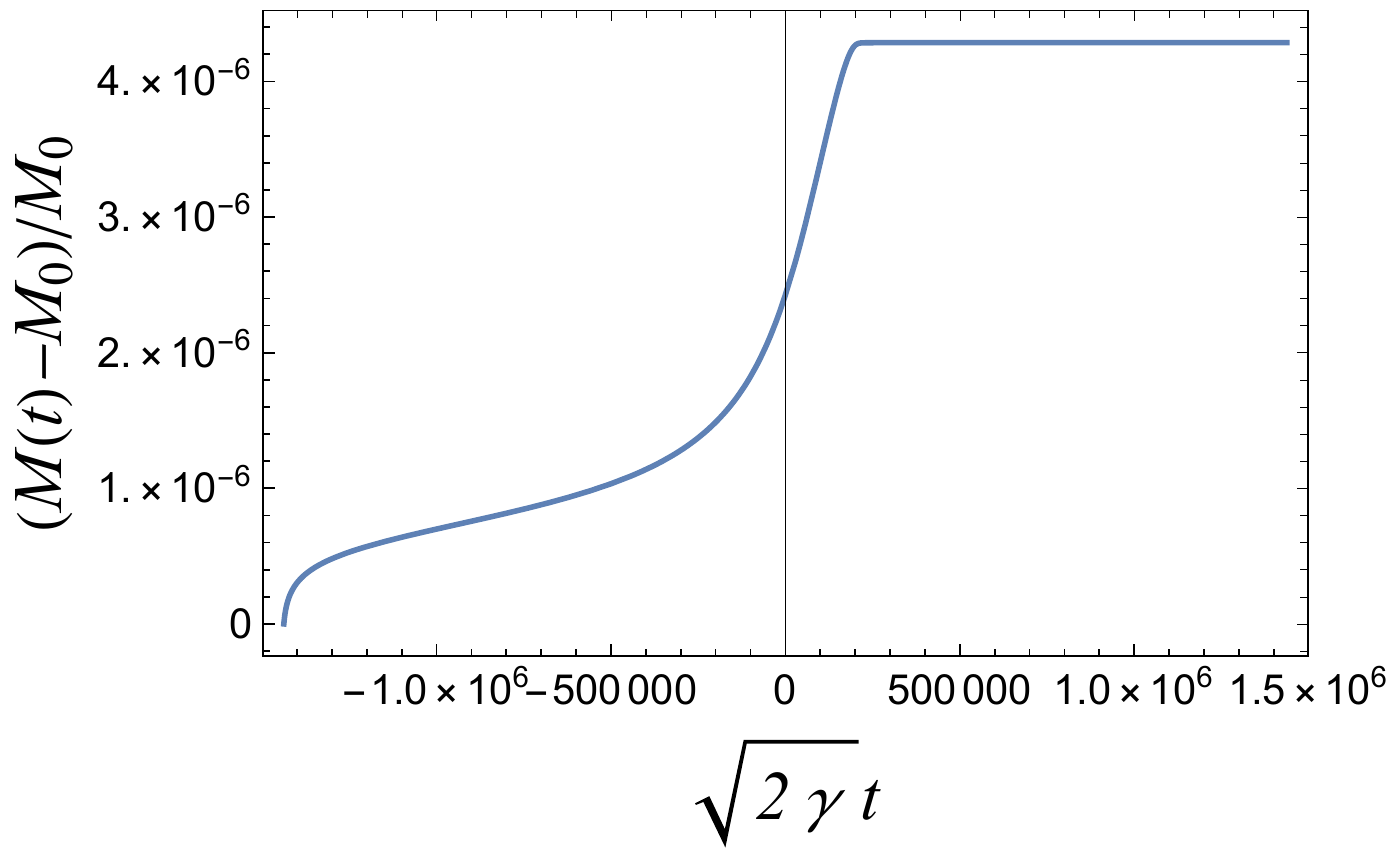}
\includegraphics[keepaspectratio, scale=0.4]{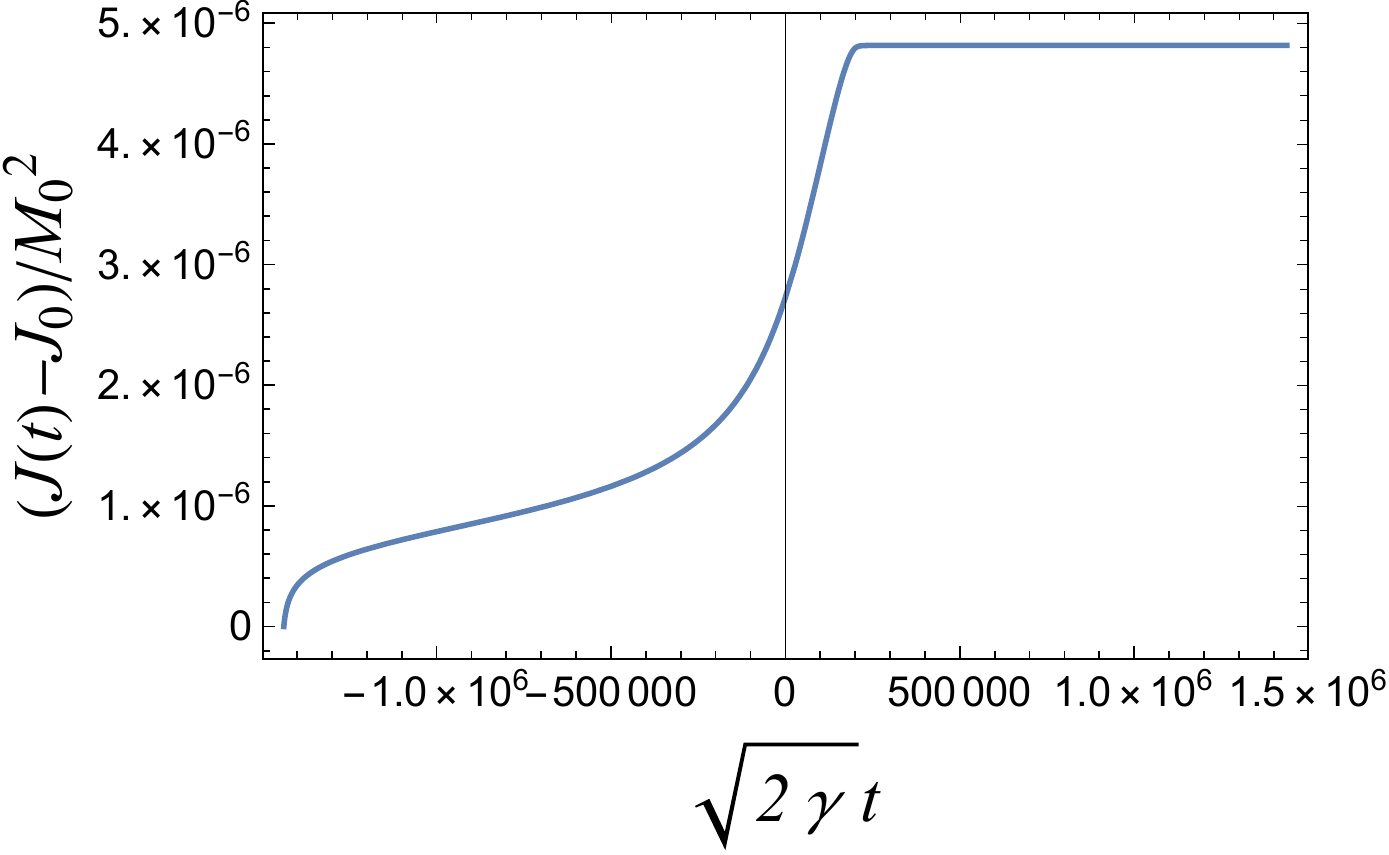}
\includegraphics[keepaspectratio, scale=0.4]{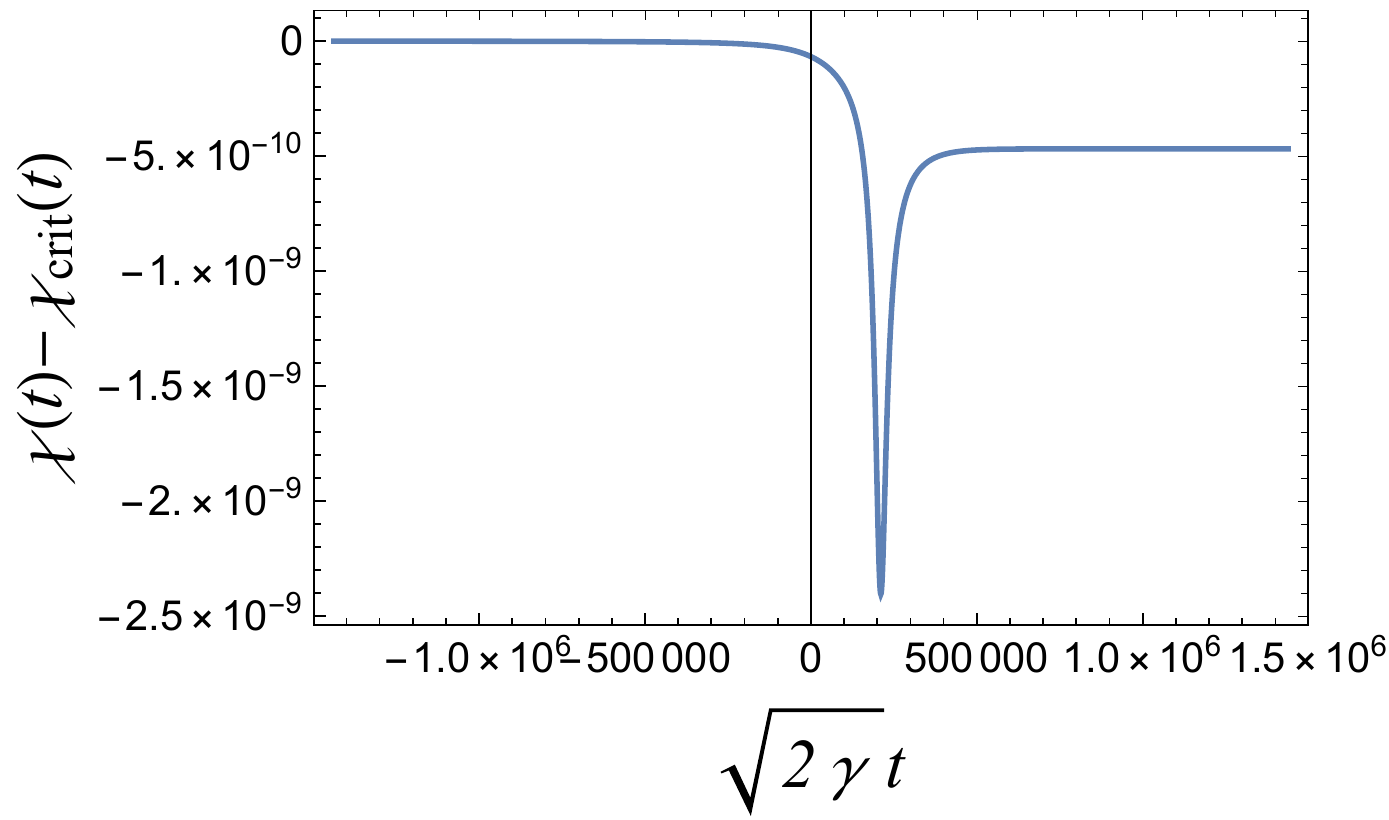}
\caption{\label{fig:a_1} Evolution of the central BH mass $M(t)$ (left), the angular momentum $J(t)$ (middle) and the deviation from the critical spin $\chi(t)-\chi_{\rm crit}$ (right) for $q=10^{-4}$, $\alpha_0=0.1$ and $M_{\rm c,0}=10^{-3}M_0$. Due to the absorption of particles belonging to the primary cloud, the mass and the angular momentum of the BH increase slightly, but it maintains the BH spin parameter slightly below the threshold value of the superradiance condition.}
\end{figure*}

First, we redefine the variables as
\begin{align}
\tilde{c}_i(t)=e^{-i\int^t dt'\left\{(\Omega-\Omega_{\rm res})-i\omega_{I}^{(1)}+i\Gamma_{\rm Gw}\right\}}c_i(t)~,
\end{align}
for $i=1,2$. Then, we can rewrite Eqs.\eqref{eq:LTc1} and \eqref{eq:LTc2} as
\begin{align}
i\frac{d\tilde{c}_i}{dt}=\sum_{j=1,2}\tilde{\cal H}_{ij}\tilde{c}_j~, \quad \tilde{\cal H}=\begin{pmatrix}
0 & \eta(t) \\
\eta(t) & \Delta(t)+i\Gamma(t)
\end{pmatrix}~, \label{eq:Sch3}
\end{align}
with
\begin{align}
\Delta(t)&=2\left(\Omega(t)-\Omega_{\rm res}(t)\right) ~,\\
\Gamma(t)&=\omega_{I}^{(2)}(t)-\omega_{I}^{(1)}(t)+\Gamma_{\rm GW}(t)~.
\label{eq:Gamma}
\end{align}
Redefined particle numbers $|\tilde{c}_i|^2$ are related to $|c_i|^2$ as
\begin{align}\label{eq:tildec}
|\tilde{c}_i(t)|^2=e^{-2\int^t dt' (\omega_{I}^{(1)}-\Gamma_{\rm GW})}|c_i(t)|^2~.
\end{align}
Now, we write
\begin{align}
\tilde{c}_2(t)=y(t)e^{-i\int^t_{-\infty} dt'(\Delta+i\Gamma)}~.
\end{align}
Substituting this into Eq.\eqref{eq:Sch3}, we have
\begin{align}\label{eq:y}
\frac{dy}{dt}=-i\eta \tilde{c}_1 e^{i\int^t_{-\infty} dt'(\Delta+i\Gamma)}~.
\end{align}
By integrating this, we formally obtain
\begin{align}
y(t)=-i\int^{t}_{-\infty}dt'\ \eta\tilde{c}_1e^{i\int^{t'}_{-\infty} dt''(\Delta+i\Gamma)}~.
\label{eq:yint}
\end{align}
If $|\Delta+i\Gamma|\gg\eta$, we can assume that the change rate of $\tilde c_1(t)$ is much slower 
than $|\Delta+i\Gamma|$. 
Then, we can carry out repeated integration by parts of the integral in Eq.~\eqref{eq:yint}, to obtain an expansion in the inverse power of $|\Delta+i\Gamma|$.
At the leading order of this expansion, we have
\begin{align}
y(t)=-\frac{\eta}{\Delta+i\Gamma}\tilde{c}_1e^{i\int^t_{-\infty} dt'(\Delta+i\Gamma)}~.
\end{align}
Then, substituting this expression for $y(t)$ into $\tilde{c}_2$ in the equation for $d\tilde c_1/dt$,  Eq.~\eqref{eq:Sch3}, and integrating it, we obtain
\begin{align}
\tilde{c}_1(t)=\exp\left(i\int^t_{-\infty}dt'\frac{\eta^2}{\Delta+i\Gamma}\right)~.
\end{align}
From the above expressions, we find that the change rate of the amplitude $\tilde{c}_1$
is much smaller than $|\Gamma|$.
Thus, from Eq.~\eqref{eq:decay_mixing}, the assumed conditions are all satisfied. 
Finally, we can write the redefined particle number for each mode as
\begin{align}
|\tilde{c}_1(t)|^2&=\exp\left(2\int^t_{-\infty}dt'\frac{\Gamma\eta^2}{\Delta^2+\Gamma^2}\right)~, \\
|\tilde{c}_2(t)|^2&=\frac{\eta^2}{\Delta^2+\Gamma^2}|\tilde{c}_1(t)|^2~.
\end{align}

Under this approximation, the equations that we need to solve 
are
Eqs.~\eqref{eq:EFbalance}, ~\eqref{eq:AMFbalance} ,~\eqref{eq:TotalAMC}, and
\begin{align}\label{eq:n1}
\frac{dn_{1}}{dt}=2\omega_{I}^{(1)}n_1+\frac{2\Gamma \eta^2}{\Delta^2+\Gamma^2}n_1-\frac{1}{M_{\rm c,0}}\frac{dE_{\rm GW}}{dt}~,
\end{align}
with
\begin{align}\label{eq:n2}
n_2=\frac{\eta^2}{\Delta^2+\Gamma^2}n_1~.
\end{align}
The last term of Eq.~\eqref{eq:n1} comes from the $i\Gamma_{\rm GW}$ in the exponential of Eq.~\eqref{eq:tildec}, and can be identified with the right-hand side of Eq.~\eqref{eq:GWn}.
In practical calculations, $\Gamma_{\rm GW}$ should be so small compared to $|\omega_{I}^{(2)}|$ that we can neglect it in $\Gamma$ (Eq.~\eqref{eq:Gamma}).
We also neglect the time derivative of $\eta$ and the higher order term of $|\Delta+i\Gamma|^{-1}$.
Now, the set of variables to be solved are $\{M,J,\Omega,n_1\}$, and we can easily solve the equations numerically for a wide range of parameters.

\section{Results}\label{Sec:results}
In this section, we show the evolution of the system obtained by solving the equations we formulated in the preceding section.
In addition, we discuss their implications for observable signatures.

\subsection{Time evolution}\label{sub:evolution}

\begin{figure}[t]
\includegraphics[scale=0.55]{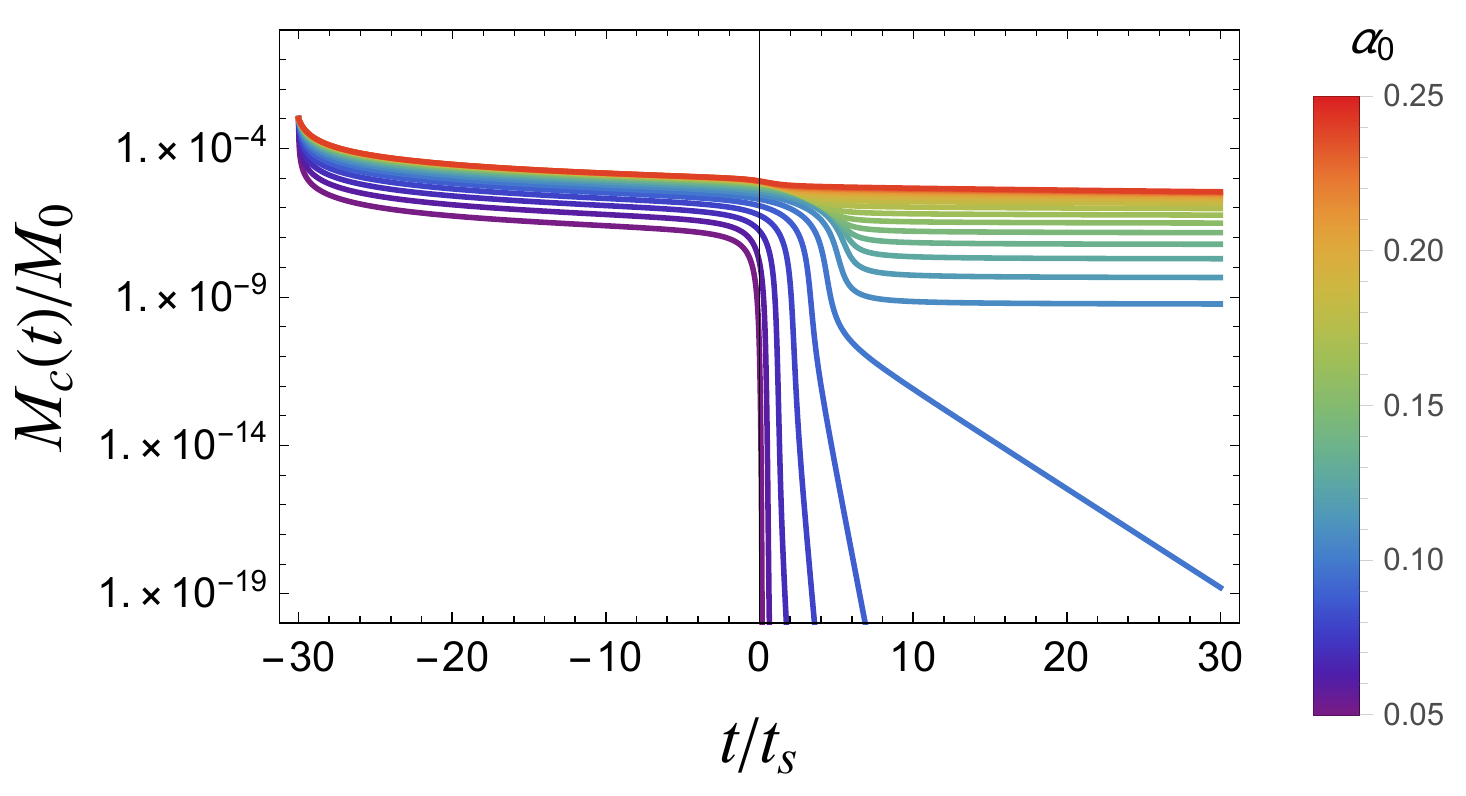}
\caption{\label{fig:alpha} Dependence of the evolution of the cloud on the gravitational fine structure constant $\alpha_0$. Each line shows the evolution of the cloud mass for $q=10^{-4}$, $M_{\rm c,0}=10^{-3}M_0$ and various $\alpha_0$. The cloud mass at a late epoch monotonically increases, as $\alpha_0$ increases.}
\end{figure}

\begin{figure}[t]
\includegraphics[scale=0.55]{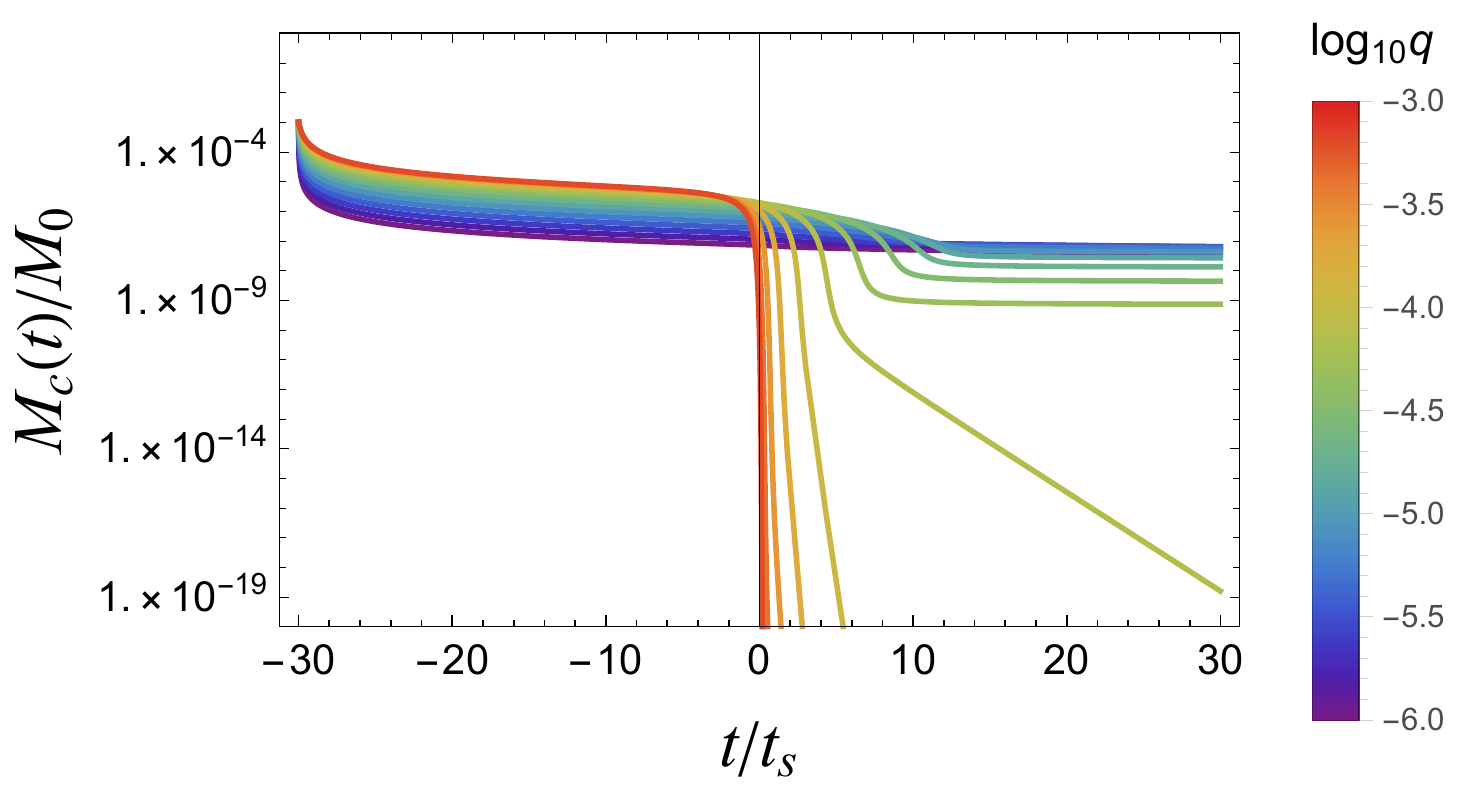}
\caption{\label{fig:q} Dependence of the cloud on the mass ratio $q$. Each line shows the evolution of the cloud mass for $\alpha_0=0.1$, $M_{\rm c,0}=10^{-3}M_0$ and various $q$. As $q$ becomes smaller, the timescale of the binary evolution becomes longer, and thus the decay due to the GW emission becomes dominant.}
\end{figure}

We first discuss the initial conditions.
To form a somewhat large cloud, the BH must have a large spin when it is formed.
However, the growth timescale of the cloud is much faster than the timescale of the binary evolution, and hence the BH spin will be quickly reduced to the threshold value for the superradiance of the dominant cloud. 
Thus, we set the initial BH spin to the threshold value, $J_0=a_{\rm crit}M_0$.
Also, how to choose the initial time is not trivial because of the decay of the cloud through the GW emission.
From the analysis of the simplified toy model in Appendix~\ref{app:toy}, we can estimate the ``start time" at which the tidal field begins to be relevant as 
\begin{align}
     t\sim-\left(1+\frac{\eta^2}{\gamma}\right)\frac{|\omega_{I}^{(2)}|}{2\gamma}\equiv -t_s\,.
     \label{def:ts}
\end{align}
We adopt $t_0=-30t_s$  evaluated with $\alpha=\alpha_0$, $\Omega=\Omega_0$ and $a=a_{\rm crit}$ as the  the initial time. 

Now, the initial condition of this system is parameterized by $\{q,\alpha_0,M_{\rm c,0}\}$.
First, let us discuss the results for the fiducial set of parameters: 
$\{q=10^{-4},\alpha_0=0.1,M_{\rm c,0}=10^{-3}M_0\}$.
The time evolution of the normalized particle number of the primary cloud $n_1(t)$ and that of the binary's orbital frequency $\Omega(t)$ are shown in Fig.~\ref{fig:n1_and_omega}.
Before reaching the resonance frequency, the particle number decreases mainly through the GW emission. 
However, since the resonance band is widened due to the presence of rapid decay of the secondary mode, characterized by $\omega_{I}^{(2)}$, the orbital frequency is slightly modified by the effect of  transition, even in this stage.

Then, when the orbital frequency gets close to the resonance frequency, the tidal interaction works more efficiently.
The particles in the first mode are transferred to the second mode, and the number $n_1$ decreases dramatically.
With the transition, the angular  momentum of the cloud is transferred to the binary orbital motion, and the orbital frequency stagnates around the resonance frequency.
Here, we should note that, because of this stagnation, the duration to pass through the resonance band becomes much longer and the net transition rate is much larger than the case when the backreaction is neglected.

After the resonance, the particle number is exponentially reduced owing to the backreaction to the central BH shown in Fig.~\ref{fig:a_1}.
Let us explain the reason why it can give such a large influence on the cloud decay after the resonance.
Initially, the superradiance condition of the primary cloud is saturated, {\it i.e.}, $\omega_{I}^{(1)}=0$.
However, once even a small number of particles are transferred to the second mode, which has an angular momentum in the opposite direction to the central BH spin, and is absorbed by the BH, the BH spin decreases slightly.
Then, the first mode becomes a non-superradiant mode, and the particles belonging to the primary cloud also begin to be absorbed by the BH.
Thus, the BH mass and angular momentum gradually increase maintaining the spin parameter slightly below the threshold value until the resonant transition becomes more efficient.

At around the peak of the resonance, the particle number of the second mode increases, and the flux to the BH of the second mode with negative angular momentum dominates that of the first mode with a positive spin.
After passing the resonance frequency, the flux of the first mode dominates again, but at that time there are
not enough particles left to spin-up the BH beyond the superradiance threshold. 
As a result, the BH spin settles to a value slightly below the threshold for the first mode to be superradiant.
Although the deviation from the critical spin is tiny, $|\omega_{I}^{(1)}|$ is sufficiently large to eliminate the cloud within the timescale of the binary inspiral. 

In summary, the cloud, first, dissipates through the GW emission.
Then, the particle number of the first mode decreases dramatically with the resonant transition, and the transferred particles to the second mode are absorbed by the BH immediately.
After that, the primary cloud decreases exponentially due to the BH spin-down below the superradiance threshold.

We also show the parameter dependence of this system.
In Fig.~\ref{fig:alpha} and Fig.~\ref{fig:q}, we show the evolution of the particle number for the same parameter but varying $\alpha_0$ and $q$, respectively.
If we neglect the backreaction and GW emission, and approximate the binary orbital frequency evolution by a linear function of $t$, the survival probability of the primary cloud is analytically evaluated as $\exp(-\pi\eta^2/\gamma)$~\cite{Baumann:2019ztm}\nobreak
\footnote{Surprisingly, this result is not changed by the presence of $\omega_{I}^{(2)}$.}.
This means that the efficiency of the tidal effect is determined by the product of the amplitude of the tidal perturbation $\eta$ and the timescale passing through the resonance band $\eta/\gamma$.
This measure of the tidal effect $\eta^2/\gamma$ is proportional to $q\alpha_0^{-11/3}$ for $q\ll1$.
Thus, the cloud mass after the resonance becomes tiny, when $\alpha_0$ is small and $q$ is somewhat large.

\subsection{Initial and final cloud mass}

\begin{figure}[t]
\includegraphics[keepaspectratio, scale=0.54]{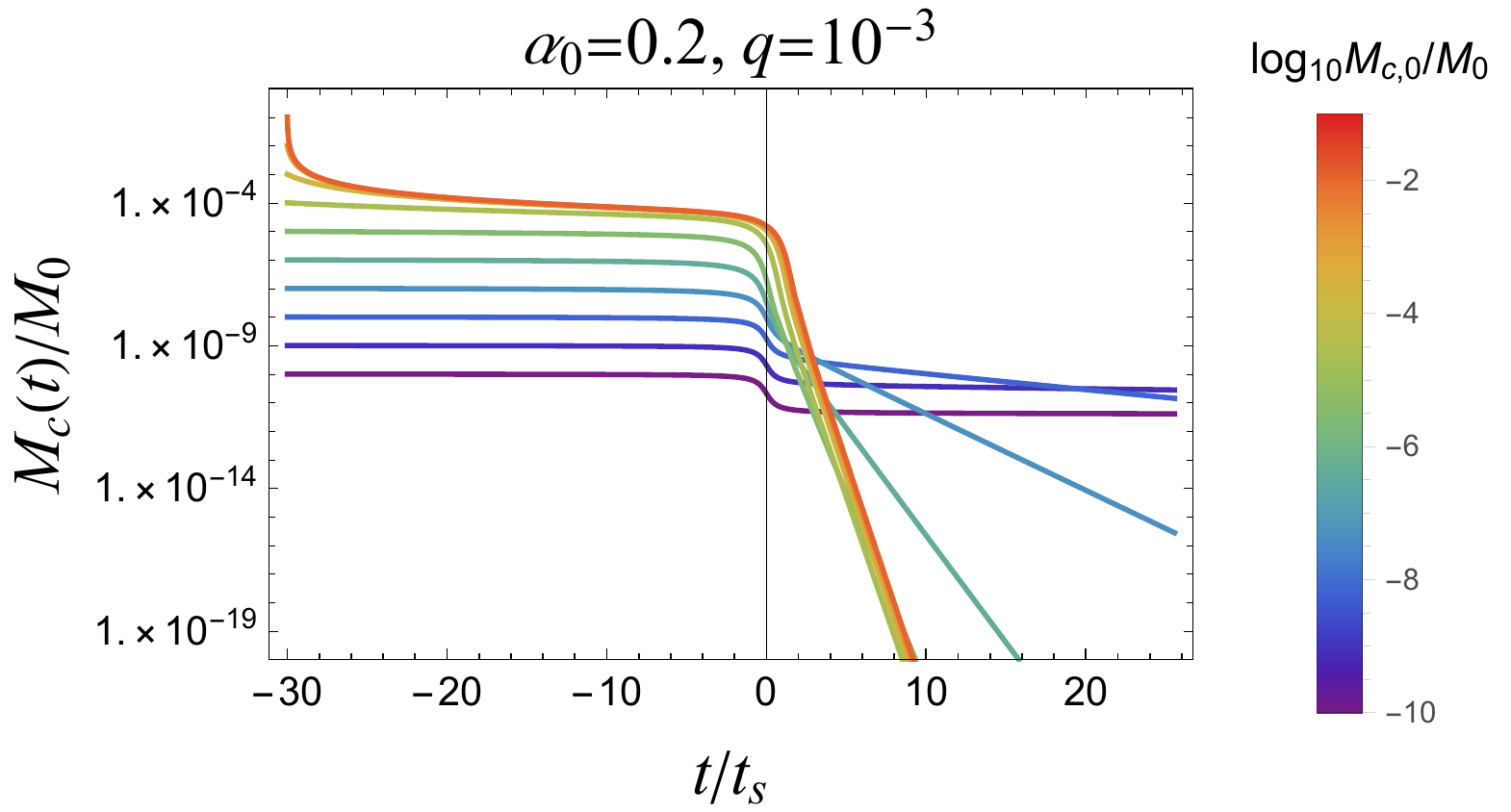} \\
\includegraphics[keepaspectratio, scale=0.54]{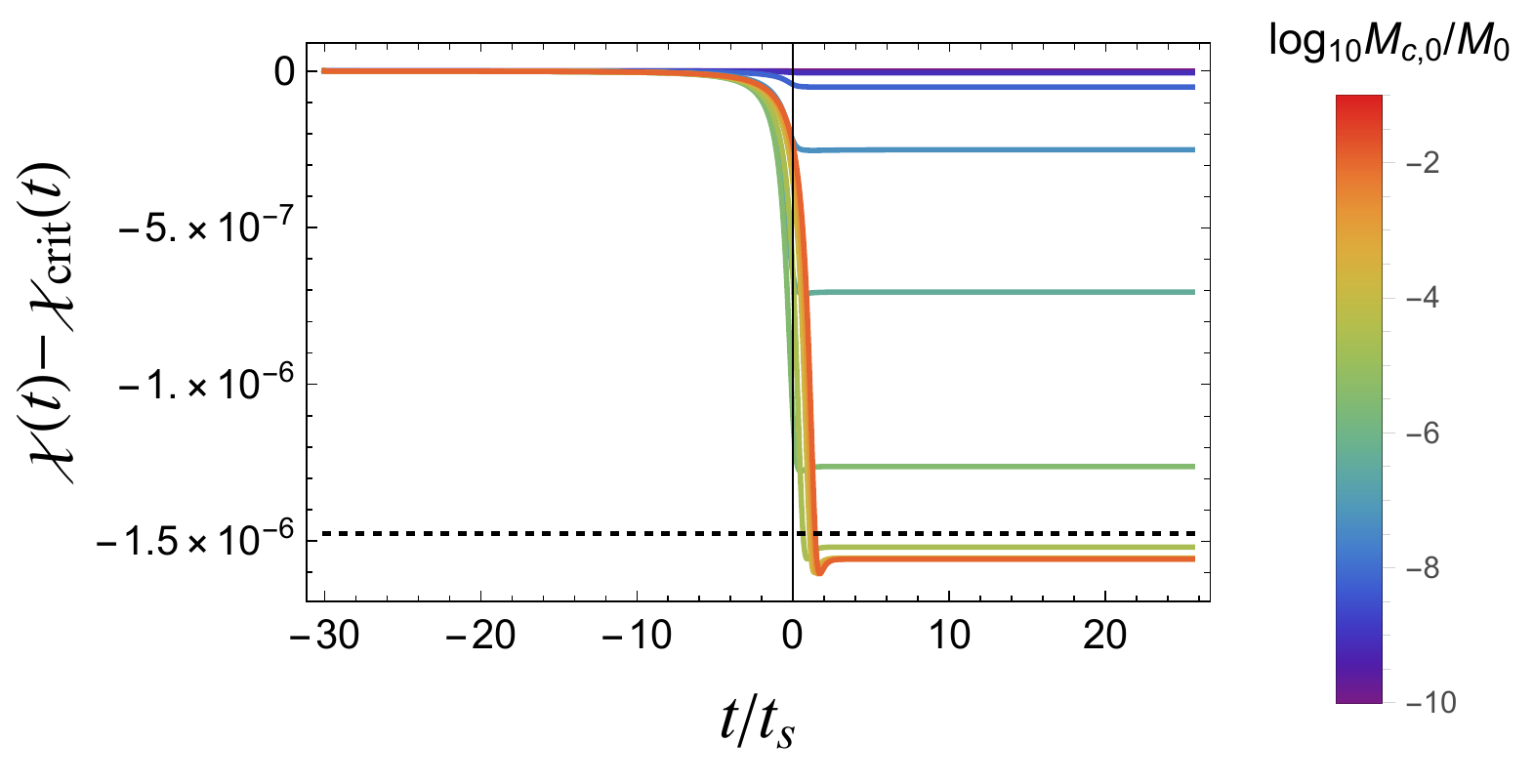}
\caption{\label{fig:case1} Case 1. Evolution of the cloud mass (top) and the BH spin (below) for $\alpha_0=0.2$, $q=10^{-3}$ and various initial cloud mass $M_{\rm c,0}$. Black dotted line in the below panel shows the minimum value of the spin estimated in Sec.~\ref{sub:amin}. In this case, the particle number after the transition is too small to spin-up the BH after the transition for a large initial cloud mass. The small initial cloud mass such that the BH spin-down is negligible gives the largest final cloud mass.}
\end{figure}

\begin{figure}[t]
\includegraphics[keepaspectratio, scale=0.54]{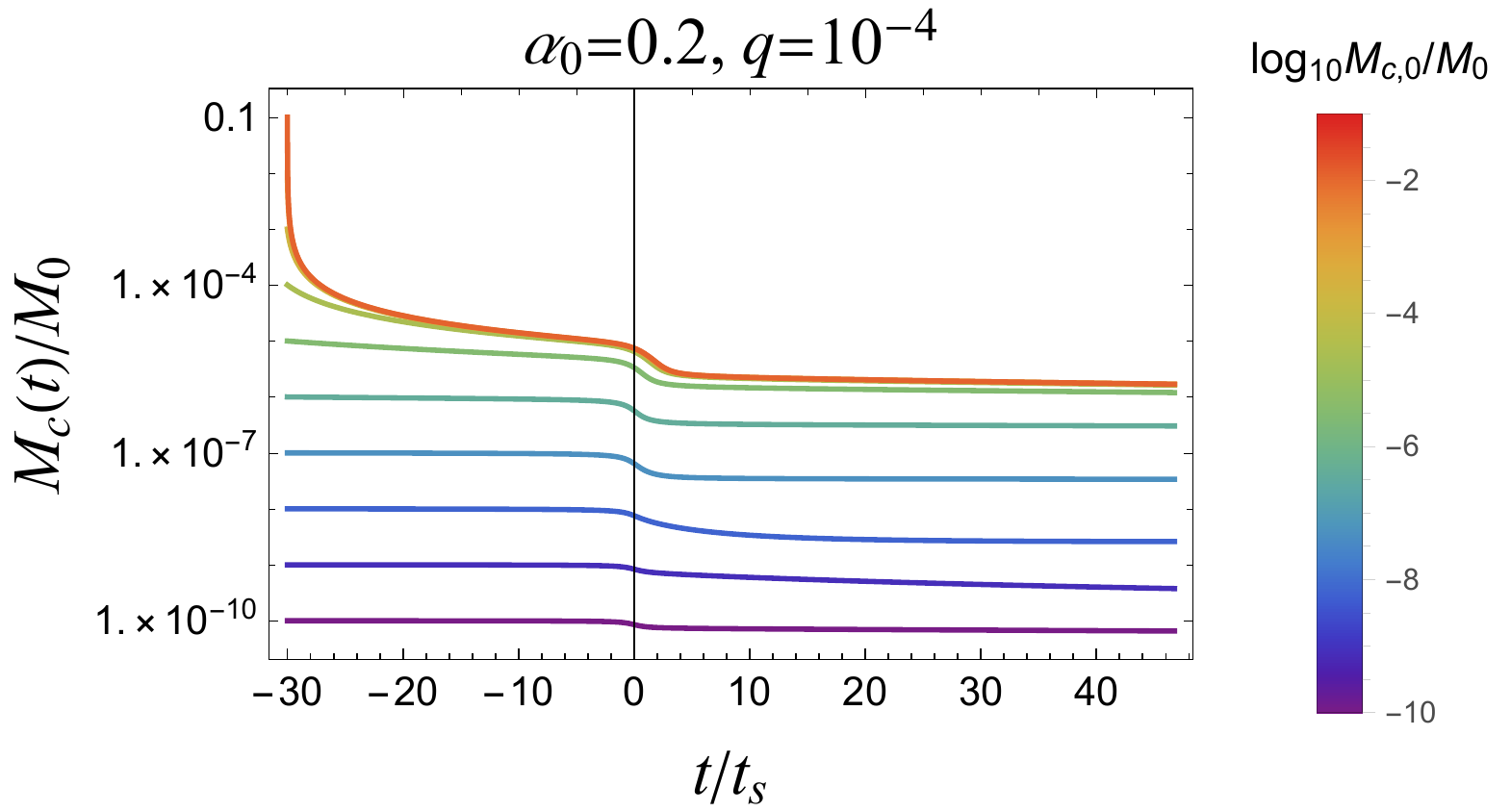} \\
\includegraphics[keepaspectratio, scale=0.54]{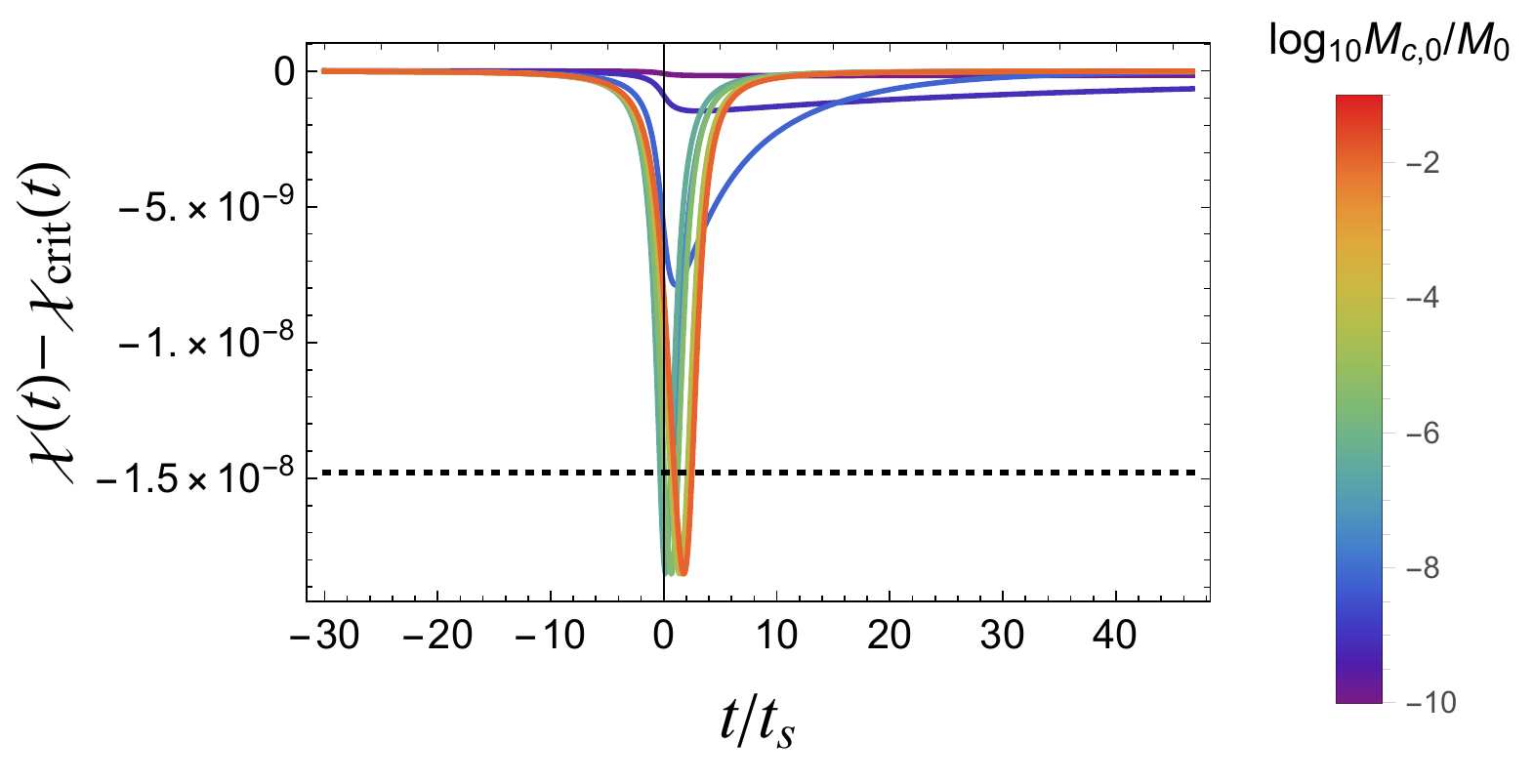}
\caption{\label{fig:case2} Case 2. The same plot as Fig.~\ref{fig:case1}, but for $q=10^{-4}$. In this case, since the transition rate is not large, there is enough particle number left to spin-up the BH after the transition for a large initial cloud mass. Thus, the cloud mass does not decrease at the late epoch, and the largest initial cloud mass gives the largest final cloud mass.}
\end{figure}

In terms of observation, it is interesting to clarify how much of the cloud can remain after the satellite passes through the resonance frequency.
The evolution of this system and the fate of the cloud also depend on the initial cloud mass.
If there are no processes that prevent the cloud's growth and the BH has nearly extremal spin when it forms, the cloud mass can be estimated as $\sim\alpha M$~\cite{Brito:2017zvb}.
In reality, however, there can be other dissipation processes besides GW emission, such as dissipation due to axion's self-interaction~\cite{Baryakhtar:2020gao,Omiya:2022gwu}. Therefore, it is worth discussing the dependence on the initial cloud mass.

We find that the initial value of the cloud mass that maximizes the final cloud mass is mainly determined by the value of the BH spin after the resonance. It can be classified into two cases, which we describe below. We show the example of the cloud mass and BH spin evolution for the initial cloud mass from $10^{-10}M_0$ to $10^{-1}M_0$ in Fig.~\ref{fig:case1} (case 1) and Fig.~\ref{fig:case2} (case 2).
Here, we take the final time as $\tau_{\rm bin}/4$, where $\tau_{\rm bin}=\Omega_0/\gamma$ is the timsescale of the binary evolution (see also Appendix~\ref{app:timescale}).

In case 1 (Fig.~\ref{fig:case1}), for large initial cloud mass, the cloud mass decreases exponentially due to the BH spin-down. In this case, since the transition rate is large, there are not enough particles left to spin-up the BH after the resonance. On the other hand, for somewhat small initial cloud mass, the particle number is too small to spin-down the BH efficiently from the beginning. In this case, the absorption to the BH can be neglected, and the final mass is determined only by the transition due to the tidal interaction.
Thus, the case with such a small initial cloud mass gives the maximum final cloud mass, for example, $M_{\rm c,0}=10^{-9}M_0$ in Fig.~\ref{fig:case1}.

In case 2 (Fig.~\ref{fig:case2}), for the largest initial cloud mass ($M_{\rm c,0}=10^{-1}M_0$),
the cloud mass does not decrease at the late epoch in this timescale.
This is because the transition rate is small and there are enough particles left to spin-up the BH by almost the threshold value of the superradiance after the resonance.
Thus, in this case, the largest initial cloud mass simply gives the maximum final cloud mass.

\begin{figure}[t]
\includegraphics[keepaspectratio, scale=0.55]{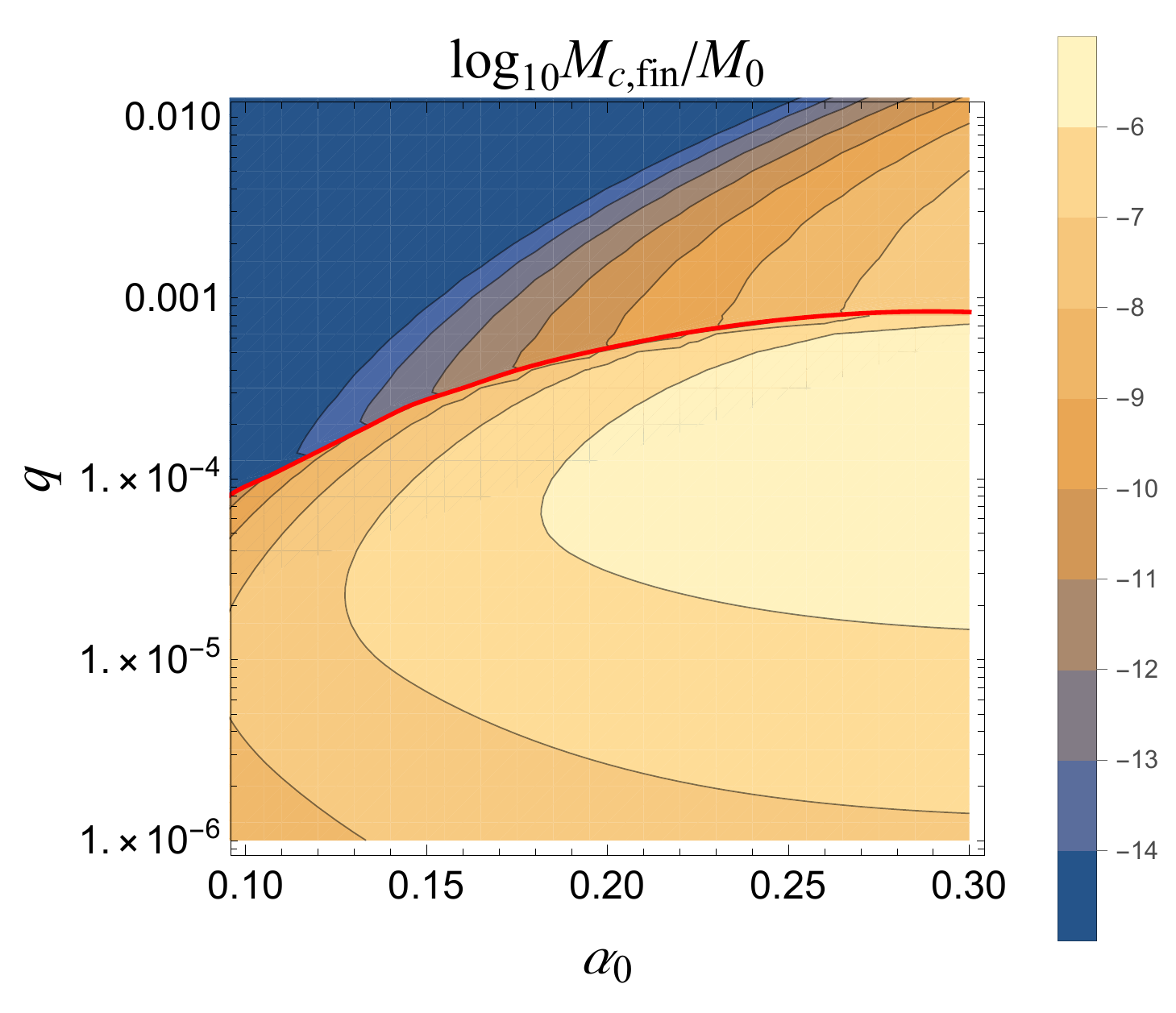}
\caption{\label{fig:maxMc} Possible maximum final mass of the cloud after the hyperfine resonance. The area above the red boundary belongs to the case1 ({\it e.g.}, Fig.~\ref{fig:case1}), and the area below it belongs to the case2 ({\it e.g.}, Fig.~\ref{fig:case2}).}
\end{figure}

We summarize the possible maximum final mass $M_{\rm c,fin}$ of the cloud after the resonance in the parameter space $(\alpha_0,q)$ in Fig.~\ref{fig:maxMc}.
We take $10^{-1}M_0$ as the largest initial cloud mass, and contours below $10^{-15}M_0$ are not shown.
The area above the red boundary belongs to the case1, and the area below it belongs to the case2.
In case1 region, the final cloud mass is mainly determined by the transition rate, {\it i.e.}, the strength of the tidal interaction characterized by $\eta^2/\gamma$. Thus, for small $\alpha_0$ and somewhat large $q$, the cloud hardly remains.
In case2 region, the final cloud mass is mainly determined by the GW emission. For small $\alpha_0$ and $q$, the timescale of the binary evolution becomes large, and thus the cloud has small mass by the time orbital frequency reaches around the resonance.
As a result, we find that the largest final mass of the cloud is $\sim 10^{-5}M_0$, which is achieved at $\alpha_0\gtrsim 0.2$ and $q\sim 10^{-3}$.

\subsection{BH spin-down}\label{sub:amin}

\begin{figure}[t]
\includegraphics[scale=0.55]{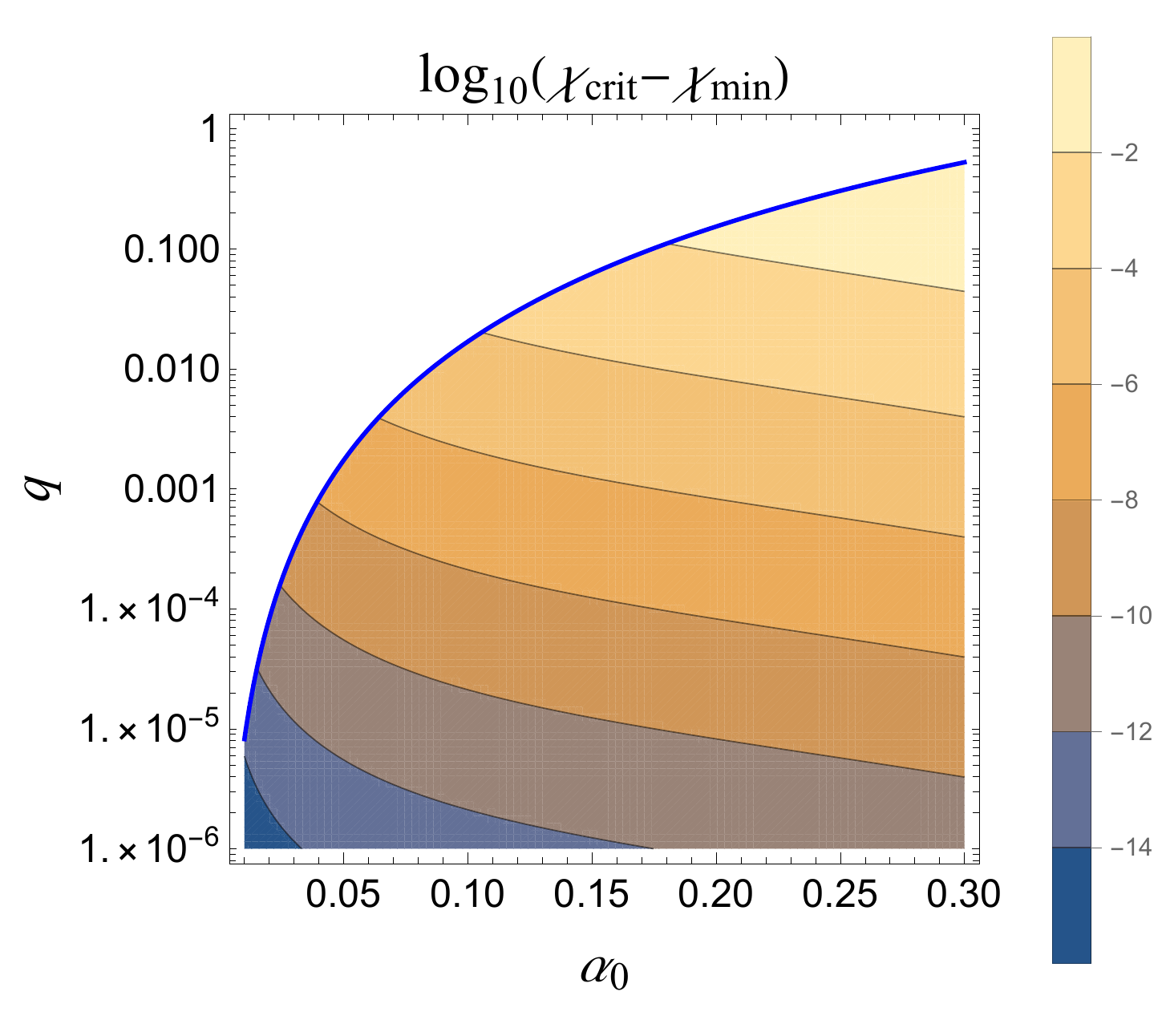}
\caption{\label{fig:amin} Approximate minimum value of the spin parameter of the central BH obtained by $d\chi/dt=0$ in Eq.~\eqref{eq:dadt}. It gives an estimation of the upper limit of the deviation from the critical spin, which can be reached by the BH spin-down. Blue solid line shows the boundary below which the hyperfine transition is relevant as Fig.~\ref{fig:region}.}
\end{figure}

In this subsection, we discuss the impact on the statistical distribution of BH spin.
If axions exist, most of BHs which experienced sufficiently large spin-up in the past
are expected to remain at the critical spin corresponding to the threshold for the superradiance (Eq.~\eqref{eq:acrit}). 
Such an accumulation of the spin distribution can be an observational signature of the existence of axions~\cite{Arvanitaki:2010sy,Brito:2014wla}. However, as we saw in the preceding subsections, axions transferred to the mode with $m=-1$ by the tidal interaction make the BH spin smaller than the critical spin.
Then, the question is, how small can the BH spin be?

To answer it, we analyze the evolution of the BH spin parameter.
From Eqs.~\eqref{eq:EFbalance} and \eqref{eq:AMFbalance}, we have
\begin{align}
    \frac{d\chi}{dt}&=-2\frac{\chi}{M}\frac{dM}{dt}+\frac{1}{M^2}\frac{dJ}{dt} \notag \\
    &=4\chi\frac{M_{\rm c,0}}{M}(\omega_{I}^{(1)}n_1+\omega_{I}^{(2)}n_2) \notag \\ & \quad +\frac{2}{\alpha}\frac{M_{\rm c,0}}{M}(-\omega_{I}^{(1)}n_1+\omega_{I}^{(2)}n_2)~. \label{eq:dadt}
\end{align}
For $\chi<\chi_{\rm crit}$ (, {\it i.e.} $\omega_{I}^{(1,2)}<0$), the first term on the right-hand side is always negative. Near the resonance, the flux of the second mode can be dominant, at which point the second term is also negative. On the other hand, when $n_2$ decreases and the flux of the first mode becomes dominant, the second term becomes positive.
Thus, we can estimate the minimum value of the BH spin parameter achieved by the reabsorption of transferred axions as $\chi_{\rm min}$ satisfying $d\chi/dt=0$ around the resonance.

Here, we use the approximation obtained in Eq.~\eqref{eq:n2} for $n_2$.
In particular, near the resonance, we can write
\begin{align}
    n_2\simeq\frac{\eta^2}{\Gamma^2}n_1~.
\end{align}
Substituting it in Eq.~\eqref{eq:dadt} and approximating $M\simeq M_0$ and $\alpha\simeq\alpha_0$, we can find the root of $d\chi/dt=0$ numerically. In Fig.~\ref{fig:amin}, we show the deviation of $\chi_{\rm min}$ obtained in this way from the critical spin $\chi_{\rm crit}$ for the parameter space $(\alpha_0,q)$.
However, it is important to stress that the deviation obtained here is only an approximate upper bound. In fact, if the cloud mass is too small, $\chi_{\rm crit}(d\chi/dt)^{-1}$ can be larger than the timescale of binary evolution as the cloud mass decreases. In that case, the BH spin-down stops before reaching the $\chi_{\rm min}$. 
In Fig.~\ref{fig:case1} and Fig.~\ref{fig:case2}, we show the evolution of the BH spin, with the dotted line corresponding to $\chi_{\rm min}$.
When the cloud mass is somewhat large, the BH spin can only go down to about $\chi_{\rm min}$ at most.
On the other hand, if the cloud mass is too small, BH spin-down terminates before reaching $\chi_{\rm min}$, and the absorption to the BH is negligible.
In particular, although it seems that the deviation of $\chi_{\rm min}$ from the critical spin for large $\alpha_0$ and small $q$ can be ${\cal O}(0.1)$ from Fig.~\ref{fig:amin}, in that region, the timescale of the binary evolution becomes small and there are no enough time to spin-down the BH.
Therefore, although the spin-down due to the absorption can be sufficiently large to deplete the cloud, it would not affect the constraints on axions from the BH spin measurements.

\subsection{Modification of the orbital frequency}

\begin{figure*}[ht]
\includegraphics[keepaspectratio, scale=0.55]{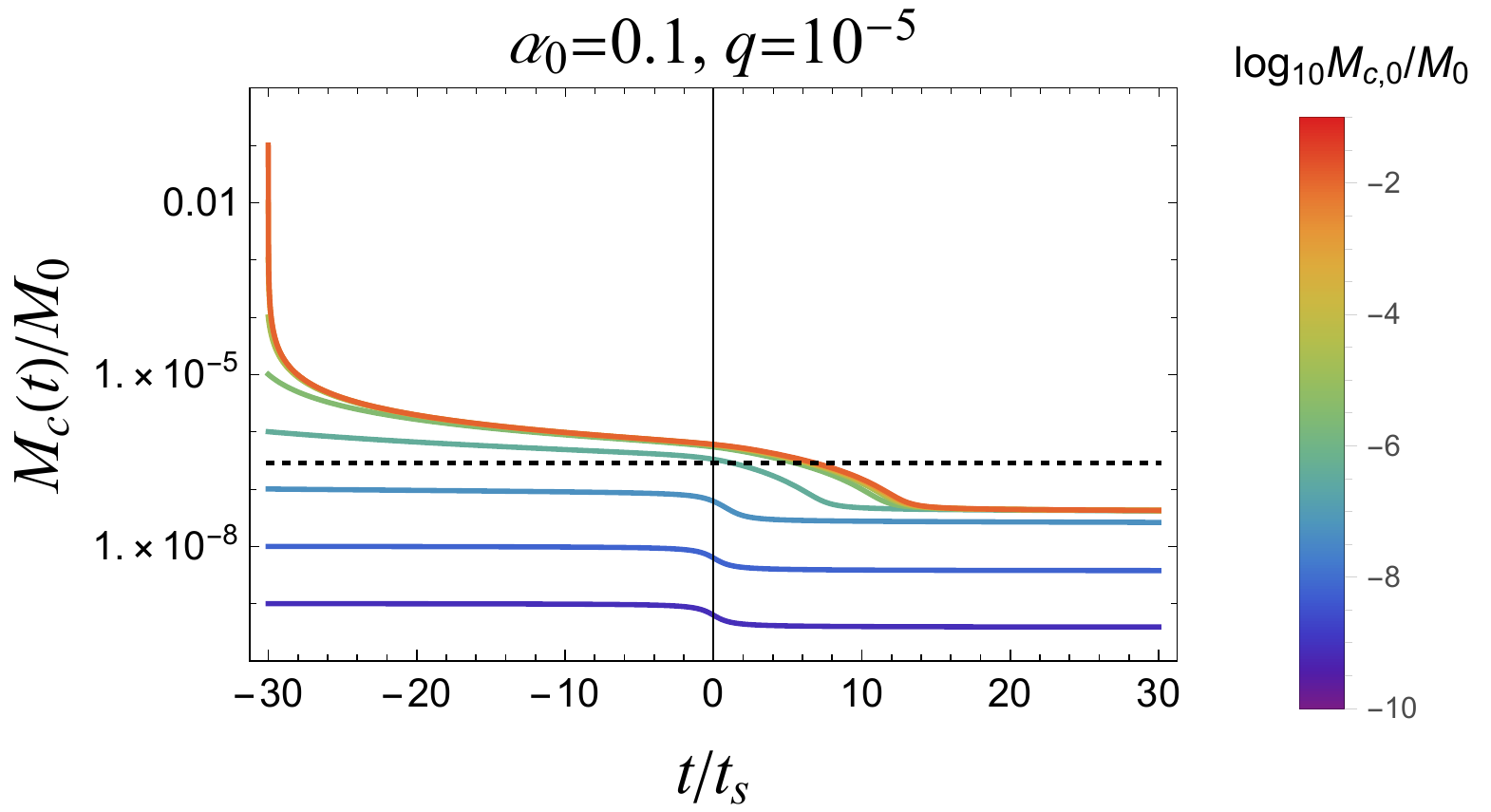}
\includegraphics[keepaspectratio, scale=0.55]{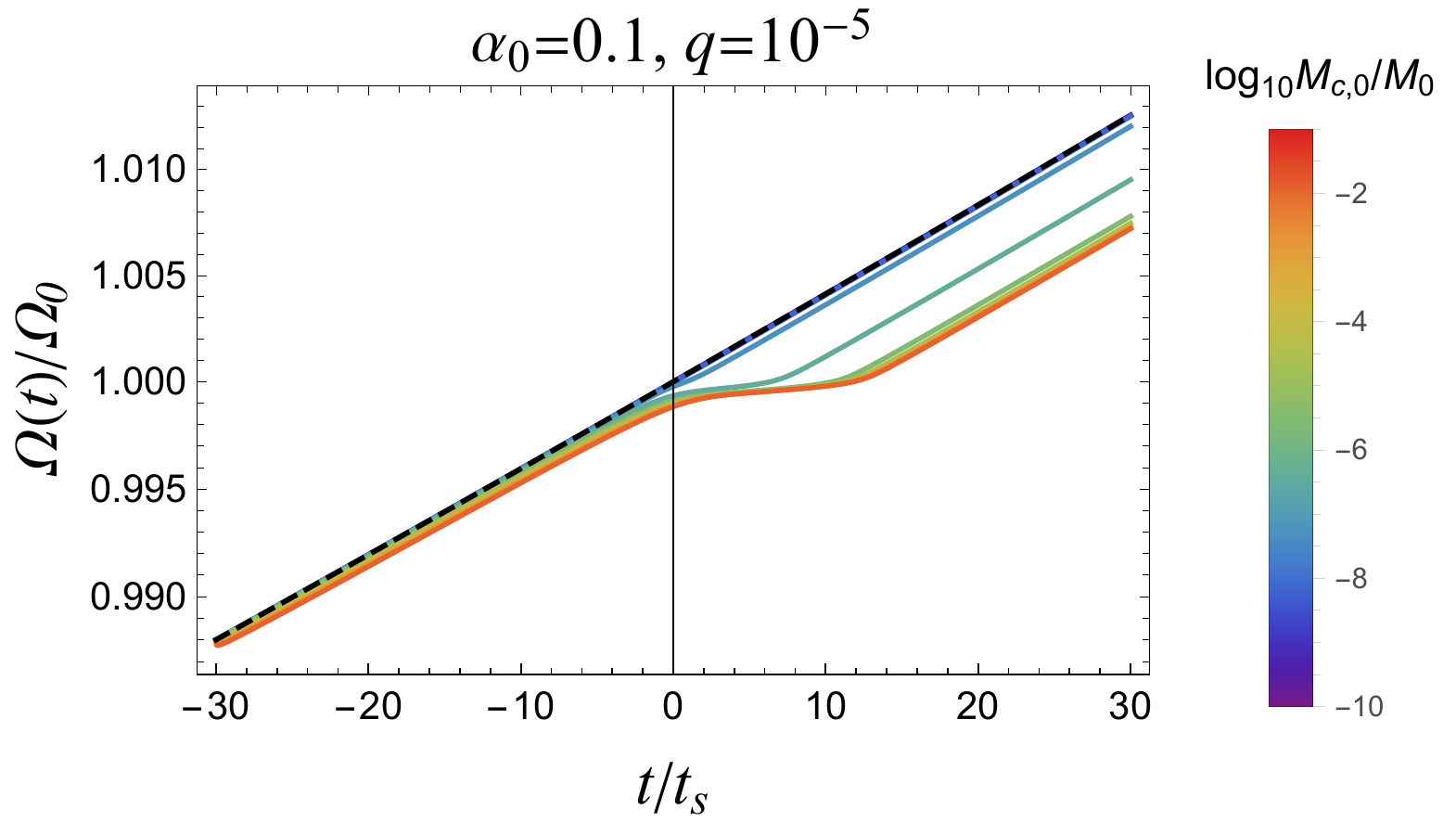}
\caption{\label{fig:float} Evolution of the cloud mass (left) and the orbital frequency (right) for $\alpha_0=0.1, q=10^{-5}$ and various initial cloud mass $M_{\rm c,0}$. Black dotted line in the left panel shows the threshold value of the cloud mass required for the backreaction to work effectively, obtained in Eq.~\eqref{eq:Mcfloat}. Black dashed line in the right panel shows the evolution of the orbital frequency in the clean binary.}
\end{figure*}

\begin{figure*}[ht]
\includegraphics[keepaspectratio, scale=0.55]{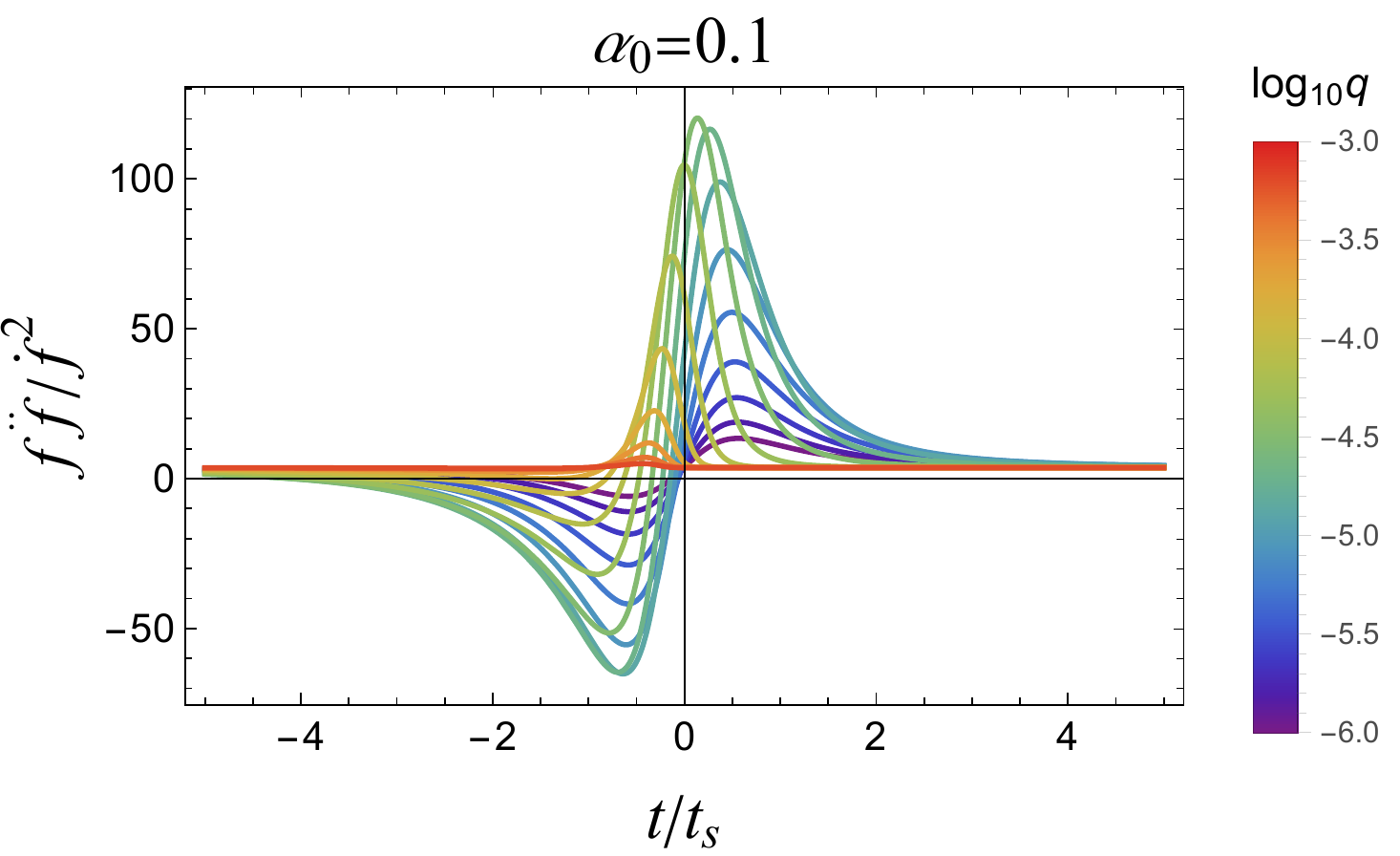}
\includegraphics[keepaspectratio, scale=0.55]{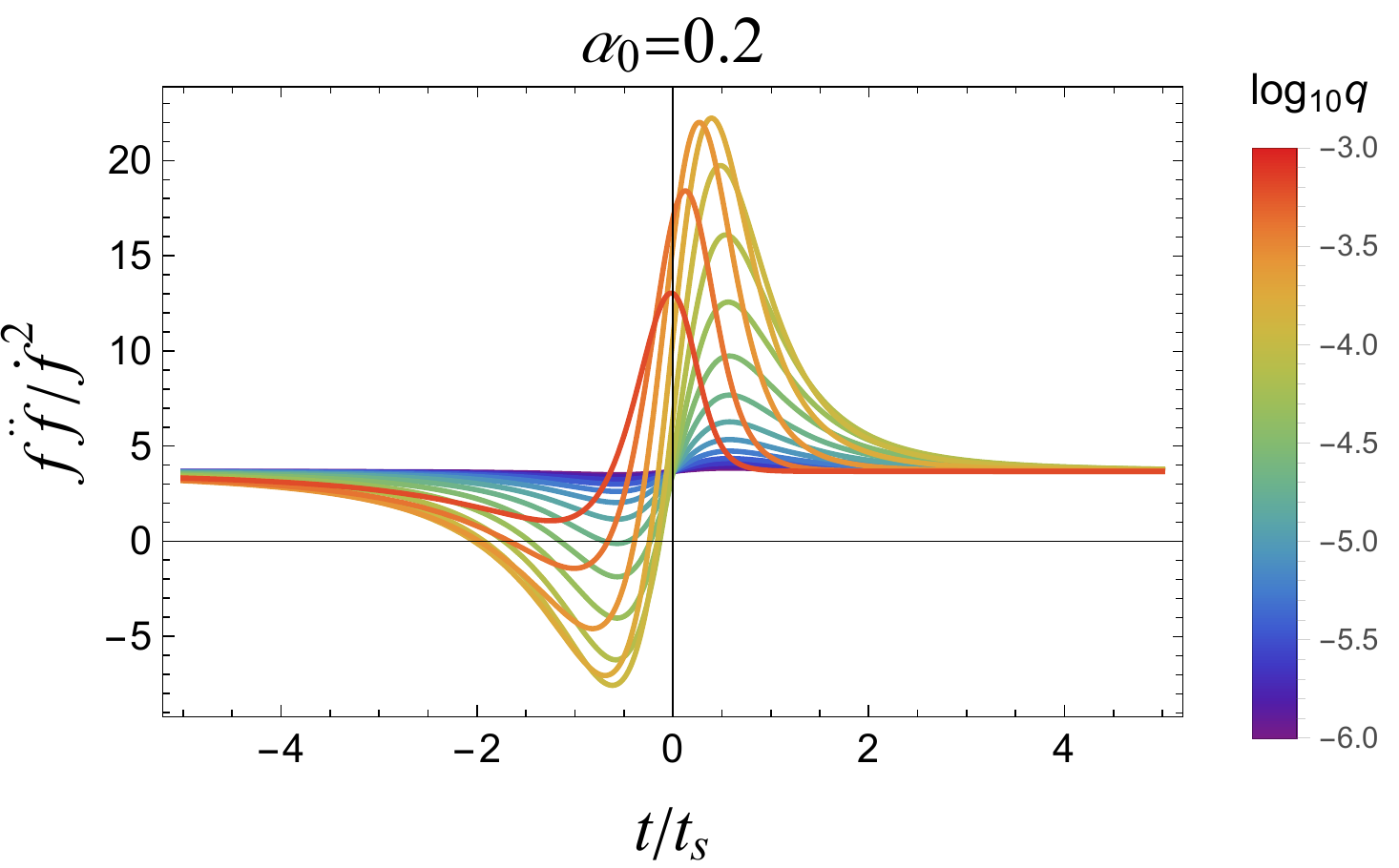}
\caption{\label{fig:fddot} Indicator in GW frequency of the presence of the cloud $f\ddot{f}/\dot{f}^2$ around the resonance $t=0$ for various $q$, $\alpha_0=0.1$ (left) and $\alpha_0=0.2$ (right). The initial cloud mass is set where the timescale of the GW emission and the binary evolution are equal, {\it i.e.}, $M_{\rm c,0}=M_{\rm c,GW}$ in Eq.~\eqref{eq:McGW}. If the binary system is clean, $f\ddot{f}/\dot{f}^2=11/3$ model-independently. Around the resonance, this quantity can be largely changed with the level transition of the cloud.}
\end{figure*}

Next, we discuss the modification of the GW frequency evolution at around the resonance.
The GW frequency at which resonance occurs is given by~\cite{Baumann:2018vus}
\begin{align}
f_{\rm res}=\frac{\Omega_{0}}{\pi}=2.2\ {\rm mHz}\ \frac{1}{1+4\alpha_0^2}\left(\frac{\alpha_0}{0.1}\right)^7\left(\frac{10M_{\odot}}{M}\right)~.
\end{align}
For typical binary systems with a supermassive BH having an extreme mass ratio companion, the resonance frequency is too low to detect.  
However, GWs at around the resonance frequency from an intermediate mass BH accompanied by a stellar mass or an even smaller mass exotic compact object could be observed by space-based GW detectors, such as LISA.

Around the resonance frequency, the orbital frequency stagnates  due to the angular momentum transfer associated with the transition. This backreaction effect also causes the delay of the rapid decrease of the cloud and enhances the transition rate.
In Fig.~\ref{fig:float}, we show the evolution of the cloud mass and the orbital frequency for $\alpha_0=0.1, q=10^{-5}$ and various initial cloud mass. 
When the cloud mass is large enough, this backreaction greatly affects the evolution.
We can estimate the threshold value of the cloud mass before the transition for the backreaction works effectively from Eq.~\eqref{eq:TotalAMC}.
For simplicity, neglecting the GW emission from the cloud and considering only the primary cloud, the orbital evolution around the resonance is approximated as
\begin{align}\label{eq:orbital:approx}
    \frac{d\Omega}{dt}\simeq\gamma+R\frac{\Omega_0}{M_0^2}\frac{dJ_{\rm c}^{(1)}}{dt}~.
\end{align}
Here, from Eq.~\eqref{eq:n1}, the time derivative of $J_{\rm c}^{(1)}$ is given by
\begin{align}
    \frac{dJ_{\rm c}^{(1)}}{dt}\simeq\frac{M_{\rm c}}{\mu}\frac{2\eta^2}{\omega_{I}^{(2)}}~.
\end{align}
For the orbital frequency to stagnate, in the right-hand side of Eq.~\eqref{eq:orbital:approx}, the first term $\gamma$ (GW radiation reaction) and the second term must be comparable. Thus, we can estimate the threshold value of the cloud mass required for the backreaction to work by equating these terms. We denote it as $M_{\rm c,float}$, and it is given as
\begin{align}
    M_{\rm c,float}&=\frac{\gamma|\omega_{I}^{(2)}|\alpha}{2R\Omega_0\eta^2}M \notag \\
    &\simeq9.5\times10^{-8}M_0(1+q)^{4/3}\left(\frac{\alpha_0}{0.1}\right)^{16/3}~.
    \label{eq:Mcfloat}
\end{align}
Therefore, even with a small mass of the cloud, we can expect that this modification can be a clear signature of the presence of an axion condensate.

Unfortunately, the timescale of the binary evolution $\tau_{\rm bin}$ is typically much longer than the observation time ($\lesssim 10\ {\rm yr}$). At first glance, it seems difficult to resolve the degeneracy with the uncertainties in the chirp mass and the mass ratio by observing the time derivatives of the GW frequency $\dot{f}$ and $\ddot{f}$.
However, we point out that $f\ddot{f}/\dot{f}^2$ can be a good indicator of deviation from clean binaries.
If the binary system is clean and the mass ratio is sufficiently small, $q\ll 1$, this non-dimensional quantity becomes a model-independent constant, {\it i.e.}, $f\ddot{f}/\dot{f}^2=11/3$ in the early stage of the inspiral.

It would be natural to assume that the cloud mass is bounded from above by the mass where the GW emission timescale $\tau_{\rm GW}$ equals the timescale of the binary evolution $\tau_{\rm bin}$ (see Appendix.~\ref{app:timescale}). Then, around the resonance,  the cloud mass, reduced only by the GW emission, is bounded by
\begin{align}
    M_{\rm c,GW}&=\frac{M^2}{\tau_{\rm bin}}\frac{\alpha^{-14}}{C} \notag \\
    &\simeq 8.9\times10^{-4}M_0\frac{q}{(1+q)^{1/3}}\left(\frac{\alpha_0}{0.1}\right)^{14/3}~. \label{eq:McGW}
\end{align}
In Fig.~\ref{fig:fddot}, we show the value of the indicator $f\ddot{f}/\dot{f}$ in the presence of a cloud with $M_{c,0}=M_{\rm c,GW}$, for $\alpha_0=0.1$ and $\alpha_0=0.2$.
They show that the deviation from clean binaries can be larger than ${\cal O}(1)$, even if the axion cloud has only a tiny fraction of the mass of the central BH.
While $q$ dependence on the indicator for the same cloud mass is weak\nobreak
\footnote{In Eq.~\eqref{eq:TotalAMC}, the main contribution to the square brackets in the second term of the right-hand side is $\dot{J}_{\rm c}^{(1)}$. From Eq.~\eqref{eq:n1}, $\dot{J}_{\rm c}^{(1)}$ is roughly proportional to $q^2$ around the resonance. Thus, $\dot{\Omega}\simeq{\cal O}(q)$. One can find that $\ddot{\Omega}\simeq{\cal O}(q^2)$ by differentiating Eq.~\eqref{eq:TotalAMC}.}, $M_{\rm c,GW}$ is approximately linearly proportional to $q$.
Thus, when the mass ratio $q$ is too small, the effect of the angular momentum transfer due to the tidal interaction also becomes small.

\section{Summary and Discussion}\label{Sec:summary}
In this paper, we have investigated the evolution of inspiralling binary systems accompanying an axion cloud before and after the orbital frequency crosses the hyperfine resonance frequency, focusing on small mass ratio ($q\ll 1$) cases.
Our main interest is how the hyperfine level transition proceeds and affects the observational signatures. 
From the comparison of timescales, we found it necessary to take into account the following components; the decaying process of the axion in the destination mode of the hyperfine transition (imaginary part of the eigenfrequency), the GW emission from the cloud, and the backreaction to the orbital motion and that to the mass and spin of the central BH.
We presented a formulation to examine the evolution of the cloud, the central BH, and the orbital motion including all these effects.
In particular, carrying out the adiabatic elimination of the degree of freedom of the amplitude of the second mode allows us to examine a wide parameter region numerically, and gives useful expressions for analyzing the behavior of the system. 

Our results show that the cloud mass is typically significantly reduced by the GW emission before the resonant transition occurs.
If $q$ is sufficiently large or $\alpha$ is sufficiently small, axions in the $m=1$ fastest growing mode are almost transferred to the $m=-1$ mode, which has angular momentum in the opposite direction to the BH spin and is easily absorbed by the BH. 
Then, the primary cloud becomes non-superradiant and can fall into the BH, 
which results in the increase of the BH angular momentum, counter-intuitively. 
However, the increase of the BH mass dominates to maintain the first mode to be non-superradiant. 
As a result, the cloud almost completely disappears by the absorption to the BH.
On the other hand, if $q$ is extremely small or $\alpha$ is sufficiently large, 
the transition rate due to tidal interaction is small. 
In such cases, since there are enough particles left to spin-up the BH again after the transition, the absorption to the BH at the late epoch can be neglected, and the cloud does not disappear completely.
However, it dissipates mainly owing to the GW emission before the transition, and the maximum mass of the cloud that can remain after the resonance is $\sim 10^{-5}M_0$ at most.
How much of axion clouds can remain after the resonance might have an implication to the survey of the cloud as an environment around the BH, such as~\cite{Bamber:2022pbs,Cole:2022fir,DeLuca:2022xlz}.

We also discussed the implication to the observational signatures.
First, we confirmed that the time variation of the BH spin around the transition is tiny, 
although this tiny variation can be important to determine the evolution of the cloud. 
This result makes robust the constraint on the existence of an axion field obtained through the BH parameter distribution measured by GWs from binary systems.
Second, we studied the influence of the transition on the inspiral GW waveform.
We found that even for extremely small cloud mass, the backreaction to the orbital motion works effectively, and the frequency stagnates around the resonance frequency.
In particular, the combination $f\ddot{f}/\dot{f}^2$ is affected by the transition to a detectable level.
Therefore, for example, the GWs from an intermediate mass BH associated with a small mass satellite can be a good target for the axion search.
We need more extensive analysis to conclude the observability of axion clouds with the modification of the waveform.
Furthermore, the generalization of the inspiral orbit and the discrimination from other environmental effects would be important. 
We leave them as future work.

\begin{acknowledgments}
T. Takahashi was supported by JST, the establishment of university fellowships towards the creation of science technology innovation, Grant Number JPMJFS2123.
TT is supported by JSPS KAKENHI Grant Number JP17H06358 (and also JP17H06357), \textit{A01: Testing gravity theories using gravitational waves}, as a part of the innovative research area, ``Gravitational wave physics and astronomy: Genesis'', and also by JP20K03928. 
%HO is supported by the Japan Society for the Promotion of Science (JSPS).
HO is supported by Grant-in-Aid for JSPS Fellows JP22J14159.
\end{acknowledgments}

\appendix

\section{Timescales}\label{app:timescale}

\begin{figure*}[ht]
\includegraphics[scale=0.7]{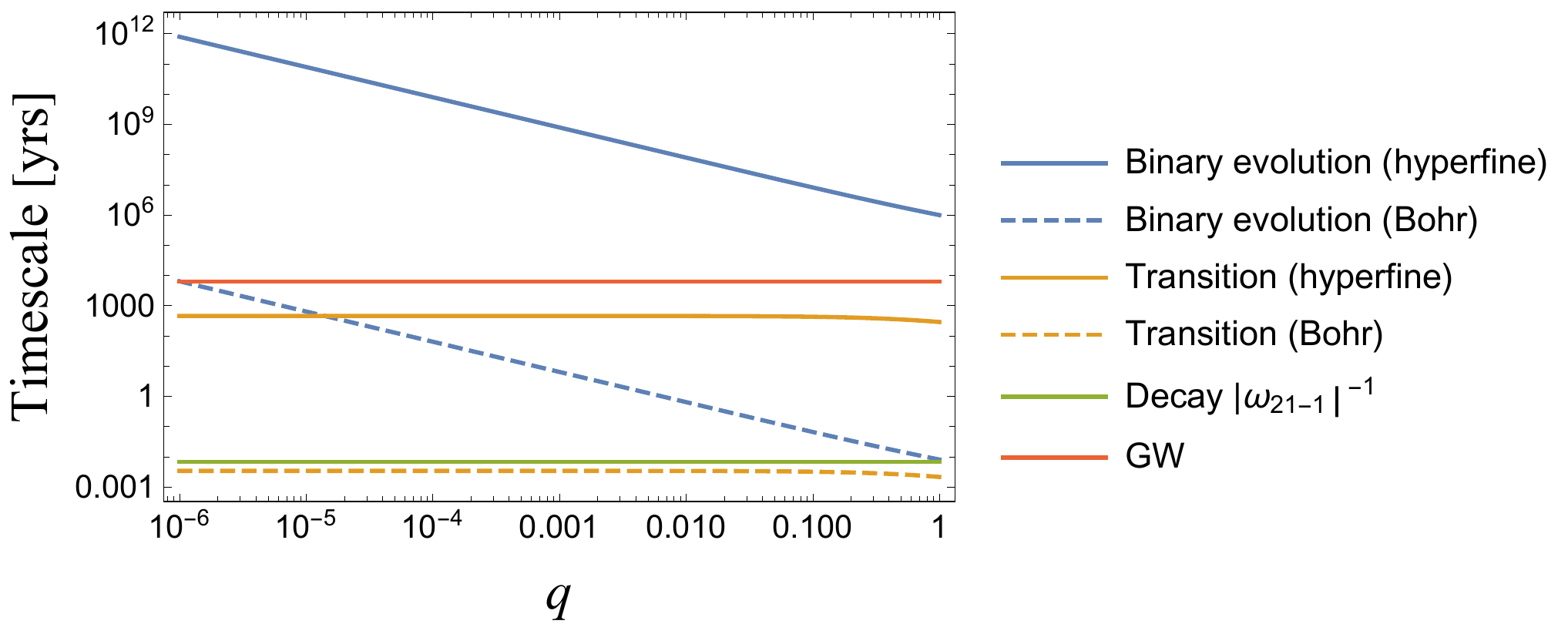}
\caption{\label{fig:timescale} Timescales involved in the resonant transition of axion clouds in binary systems for $\alpha=0.1$ and $M=M_{\odot}$. Blue and yellow solid lines show the timescale of the binary evolution and the transition at the hyperfine resonance, respectively. Blue and yellow dashed lines show the same quantities, but for the typical Bohr transition ($\ket{211}\to\ket{31-1}$). Green and red lines show the timescales of decay of the secondary cloud ($\ket{21-1}$) and of the GW emission of the primary cloud ($\ket{211}$) for $M_{\rm c,0}=0.1M$, respectively.}
\end{figure*}

\begin{table*}
\caption{\label{tab:timescale}Timescales involved in the hyperfine resonance of axion clouds.}
\begin{tabular}{|c|c|}
\hline
 process & time \\ \hline \vspace{-10pt}& \\  \hline
 Binary evolution & $\displaystyle \tau_{\rm bin}=2.2\times10^{13}\ {\rm s}\frac{(1+q)^{1/3}}{q}\left(\frac{M}{M_{\odot}}\right)\left(\frac{\chi}{0.4}\right)^{-8/3}\left(\frac{\alpha}{0.1}\right)^{-16}$ \\ \hline
 Transition & $\displaystyle \tau_{\rm trans}=1.3\times10^{10}\ {\rm s}\frac{1}{(1+q)^{2/3}}\left(\frac{M}{M_{\odot}}\right)\left(\frac{\chi}{0.4}\right)^{-5/3}\left(\frac{\alpha}{0.1}\right)^{-13}$ \\ \hline
 Decay of the secondary mode & $\displaystyle \tau_{\rm inst}\simeq5.9\times10^{5}\ {\rm s}\left(\frac{M}{M_{\odot}}\right)\left(\frac{\chi}{0.4}\right)^{-1}\left(\frac{\alpha}{0.1}\right)^{-9} $ \\ \hline
 GW emission & $\displaystyle \tau_{\rm GW}=2.0\times10^{11}\ {\rm s}\left(\frac{M}{M_{\odot}}\right)\left(\frac{M_{\rm c,0}/M}{0.1}\right)^{-1}\left(\frac{\alpha}{0.1}\right)^{-14}$ \\ \hline
\end{tabular}
\end{table*}

In this appendix, we summarize the timescales involved in our problem.

\noindent
%Light crossing time:
%\begin{align}
%\tau_{\rm BH}=M~.
%\end{align}
\begin{description}
\item[Binary evolution]\mbox{}\\ The timescale of the binary evolution due to the GW radiation at the resonance frequency $\Omega_0$ is given by
\begin{align}
\tau_{\rm bin}=\frac{\Omega_0}{\gamma}=\frac{5}{96}M\frac{(1+q)^{1/3}}{q}(M\Omega_0)^{-8/3}~,
\end{align}
where $\gamma$ is defined by Eq.~\eqref{def:gamma}.
\item[Transition]\mbox{}\\
The resonance bandwidth can be estimated as $\Delta\Omega\sim2\eta$. Hence, if one can neglect the instability of the mode of the transition destination and linearize the orbital evolution, the timescale for passing through the resonance band is given by
\begin{align}
\tau_{\rm trans}=\frac{2\eta}{\gamma}~.
\end{align}
\item[Decay of the secondary cloud]\mbox{}\\
The secondary cloud decreases as $\sim e^{-2|\omega_{I}^{(2)}|t}$, and the timescale is given by
\begin{align}
\tau_{\rm inst}=|\omega_{I}^{(2)}|^{-1}~.
\end{align}
\item[GW emission of the primary cloud]\mbox{}\\
From the energy conservation $\dot{M}_{\rm c}=-\dot{E}_{\rm GW}$ (see Eq.~\eqref{GWflux}), one can obtain
\begin{align}
M_{\rm c}(t)=\frac{M_{\rm c,0}}{1+(t-t_0)/\tau_{\rm GW}}~.
\end{align}
Here, the timescale is given by
\begin{align}
\tau_{\rm GW}=\frac{1}{C}\frac{M^2}{M_{\rm c,0}}\alpha^{-14}~.
\end{align}
\end{description}

Parameter dependencies of the timescales mentioned above are summarized in Fig.~\ref{fig:timescale} and Table~\ref{tab:timescale}. %Note that the meaning of ``timescale" is different for each, but they give a criterion for consideration.

\section{Toy model for adiabatic elimination}\label{app:toy}

\begin{figure}[t]
\begin{tabular}{c}
\begin{minipage}[t]{1\hsize}
\centering
\includegraphics[scale=0.65]{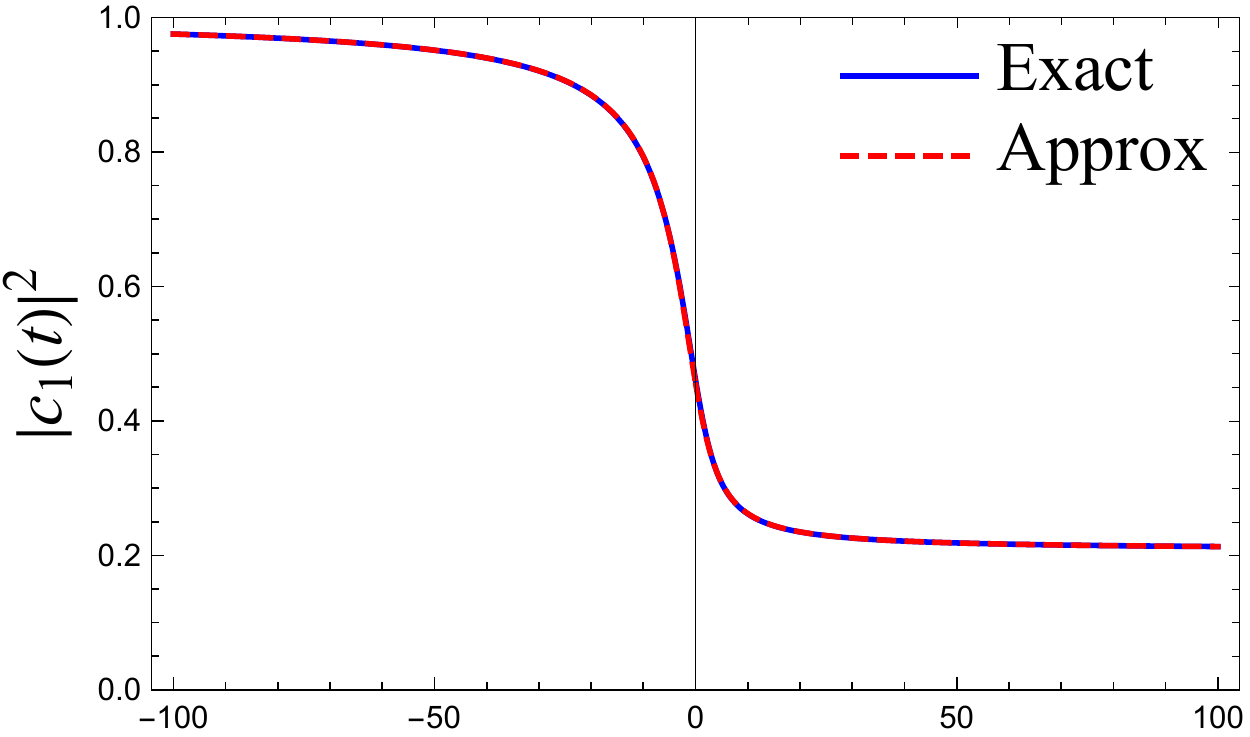}
\end{minipage} \\
\begin{minipage}[t]{1\hsize}
\centering
\includegraphics[scale=0.65]{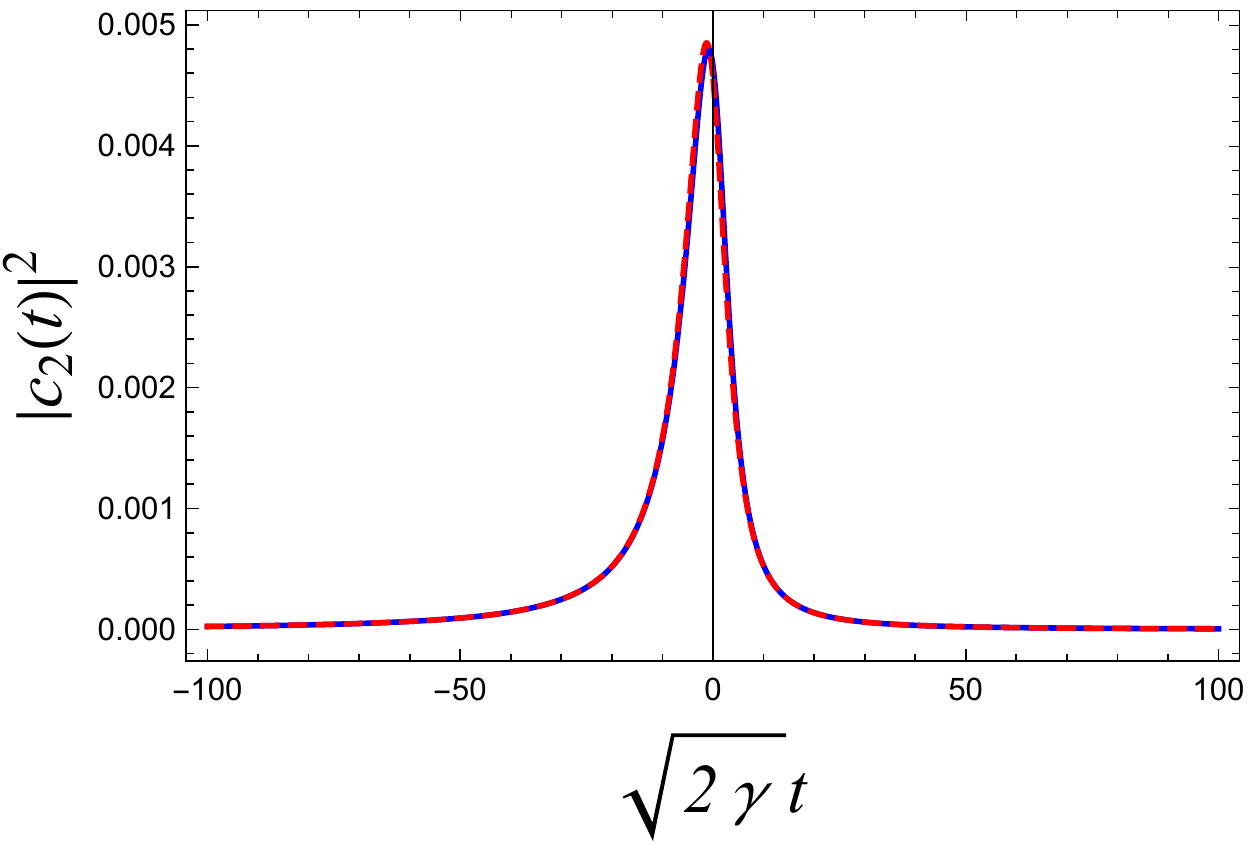}
\caption{\label{fig:app} Time evolution of the particle number of each mode for $\eta/\sqrt{2\gamma}=0.5$ and $\omega_{I}/\sqrt{2\gamma}=5$, {\it i.e.}, $\omega_{I}/\eta=10$. Blue solid line shows the exact solution and red dashed line shows the approximate solution obtained by the adiabatic elimination.}
\end{minipage}
\end{tabular}
\end{figure}

In this appendix, we discuss the approximation used in Sec.~\ref{sub:elimination} with a simplified toy model. Consider the two level transition described by the Schr$\ddot{\rm o}$dinger equation
\begin{align}
i\frac{d}{dt}\begin{pmatrix}c_1 \\ c_2 \end{pmatrix}=\begin{pmatrix}
0 & \eta \\
\eta & \Delta(t)-i\omega_{I}
\end{pmatrix}
\begin{pmatrix}c_1 \\ c_2 \end{pmatrix}~.
\end{align}
Let $\eta$ and $\omega_{I}$ be constants and $\Delta(t)=2\gamma t$ ($\gamma$ is constant).
This model is a simplification of ignoring all backreactions, GW emissions and linearizing the binary evolution in the problem we investigate.
If $\omega_{I}=0$, this model is known as Landau-Zener problem~\cite{Baumann:2019ztm,10011873546,Zener:1932ws}.
Now, we want to study the level transition to the decaying mode ($\omega_{I}>0$).
For this problem, we have an exact analytic solution with the initial conditions $c_1(-\infty)=1$ and $c_2(-\infty)=0$ as~\cite{PhysRevA.46.4110,PhysRevA.55.2982}
\begin{widetext}
\begin{align}
|c_1(t)|^2&=e^{-\omega_{I} t-\frac{\pi}{2}\frac{\eta^2}{2\gamma}}\left|D_{i\eta^2/2\gamma}\left(e^{i\frac{3\pi}{4}}(\sqrt{2\gamma}t-i\omega_{I}/\sqrt{2\gamma})\right)\right|^2~, \\
|c_2(t)|^2&=e^{-\omega_{I} t-\frac{\pi}{2}\frac{\eta^2}{2\gamma}}\frac{\eta^2}{2\gamma}\left|D_{i\eta^2/2\gamma-1}\left(e^{i\frac{3\pi}{4}}(\sqrt{2\gamma}t-i\omega_{I}/\sqrt{2\gamma})\right)\right|^2~,
\end{align}
\end{widetext}
where $D_\nu(z)$ is the parabolic cylinder function.

Carrying out the adiabatic elimination as Sec.~\ref{sub:elimination}, we obtain the approximate solution for the particle number as
\begin{align}\label{eq:toyn1}
|c_1(t)|^2&\simeq\exp\left(2\int_{-\infty}^{t}dt'\frac{\omega_{I}\eta}{4\gamma^2t^{'2}+\omega_{I}^{2}}\right) \notag \\
&=\exp\left[-\frac{\eta^2}{\gamma}\left(\arctan\frac{2\gamma t}{\omega_{I}}+\frac{\pi}{2}\right)\right]~,
\end{align}
and
\begin{align}
|c_2(t)|^2 \simeq \frac{\eta^2}{4\gamma^2t^2+\omega_{I}^{2}}|c_1(t)|^2~.
\end{align}
In Fig.~\ref{fig:app}, we compare the approximate solution obtained by the adiabatic elimination with the exact one. As one can confirm from the figure, the two solutions agree quite well when $\omega_{I}/\eta$ is sufficiently large.
 
We can estimate the time when the perturbation starts to work from Eq.~\eqref{eq:toyn1}.
For $|t|\gg\omega_{I}/2\gamma\ (t<0)$, one can expand $|c_1(t)|^2$ with respect to $1/|t|$ as
\begin{align}
|c_1(t)|^2\sim\exp\left(-\frac{\eta^2 \omega_{I}}
{2\gamma^2|t|}\right)~.
\end{align}
If $\eta^2/\gamma\gg1$, the exponent can be ${\cal O}(1)$,
even for $|t|\gg\omega_{I}/2\gamma$. In this case, the proper choice of the time for the onset of the perturbation would be $t\sim-\eta^2\omega_{I}/2\gamma^2$.
On the other hand, if $\eta^2/\gamma\lesssim1$, the exponent in Eq.~\eqref{eq:toyn1}
vanishes for $|t|\gg\omega_{I}/2\gamma$. 
In this case, it is enough to choose the starting time at $t\sim-\omega_{I}/2\gamma$.
Combining them, we have the estimation of the time when the perturbation starts to work as $t\sim-(1+\eta^2/\gamma)(\omega_{I}/2\gamma)$.

\bibliography{ref}

\end{document}